\documentstyle[preprint,version2,aps]{revtex}
\def\sla{\raise.15ex\hbox{$/$}\kern-0.57em}
\begin{document}
\draft
\preprint{OCIP/C 93-18}
\preprint{UQAM-PHE-9306}
\preprint{hep-ph/9401306}
\preprint{December 1993}
\begin{title}
A PHENOMENOLOGICAL STUDY OF THE PROCESS \\
 $e^+e^-\to\mu^+\mu^-\nu_l\bar\nu_l$ AT HIGH ENERGY $e^+e^-$ COLLIDERS \\
AND MEASUREMENT OF THE $ZWW$ AND $\gamma WW$ COUPLINGS
\end{title}
\author{Gilles Couture\footnote{e-mail: couture@osiris.phy.uqam.ca}}
\begin{instit}
D\'epartement de  Physique, Universit\'e du Qu\'ebec \`a Montr\'eal \\
C.P. 8888, Succ. A, Montr\'eal, Qu\'ebec, Canada, H3C 3P8
\end{instit}
\moreauthors{Stephen Godfrey\footnote{e-mail: godfrey@physics.carleton.ca}}
\begin{instit}
Ottawa-Carleton Institute for Physics \\
Department of Physics, Carleton University, Ottawa CANADA, K1S 5B6
\end{instit}
\begin{abstract}
We perform a detailed study of the process
$e^+e^-\to \mu^+\mu^-\nu_l\bar\nu_l$ including all contributions.
The contributions other than from real gauge boson production leads to a
rich phenomenology.
We explore the use of the process as a means of precision
measurement of the $ZWW$ and $\gamma WW$ vertices.
We concentrate on
LEP II energies, $\sqrt{s}=200$ GeV, and energies appropriate to the
proposed Next Linear Collider (NLC) high energy $e^+e^-$ collider
with center of mass energies $\sqrt{s}=500$
and 1~TeV. At 200 GeV, the process
offers, at best, a consistency check of other processes
being considered at LEP200.  At 500~GeV,
the parameters $\kappa_\gamma$, $\lambda_\gamma$, $\kappa_Z$,
and $\lambda_Z$ can be measured to about $\pm 0.1$ or better at 95\% C.L.
while at 1 TeV, they can be measured to about $\pm 0.01$.  At the
high luminosities anticipated at high energy linear colliders precision
measurements are likely to be limited by systematic rather than statistical
errors.
\end{abstract}

\narrowtext
\section{INTRODUCTION}
\label{sec:intro}

There is a growing interest in the physics that can be studied at high energy
$e^+e^-$ colliders \cite{nlc}.  High energy $e^+e^-$ colliders offer a
cleaner environment than multi-TeV hadron colliders and are
therefore expected to allow more quantitative studies of
physics at the Fermi scale.  Some of the physics topics that
have been explored are precision measurement of $t$-quark
properties,  searches for new physics,  electroweak symmetry
breaking,  tests of QCD, and precision measurements of the
electroweak gauge bosons \cite{nlcphysics}.

At the same time there is a growing appreciation that to
realistically assess the physics potential of a specific
process one must perform detailed studies of the final state decay products
that will be observed by a detector rather than the massive,
short lived states that we are directly interested
in\cite{fourfermi,aeppli,couture,kalyniak,eptoz1,yehudai}.
Performing such a study
greatly increases the complexity of the analysis as one must include finite
width effects of the decaying particles and all the background processes
that result in the same final state.  On the other hand, this complexity
results in a much richer phenomenology which more closely describes what is
experimentally observed.  In addition, the finite width effects are, in some
sense,  radiative corrections of order $\Gamma/M \sim \alpha$ which must
ultimately be included in a full calculation including radiative corrections
\cite{aeppli}.

In this paper we present a detailed study of the process $e^+e^- \to
\nu\bar{\nu} \mu^+ \mu^-$ motivated by our interest in the underlying process
$e^+e^- \to \nu_e \bar{\nu}_e Z^0$.  Although this process has been studied
elsewhere \cite{hagiwara,ambrosanio}, none of the
previous calculations have included the decay to final state fermions with
finite width effects and the nonresonant backgrounds.  We find that
including these contributions adds considerably to the richness of the
phenomenology.  We then use this
process to study the $WW\gamma$ and $WWZ^0$ couplings.

Although experiments at the CERN LEP-100 $e^+e^-$ collider
and the SLAC SLC $e^+e^-$ collider\cite{lep} have provided
stringent tests \cite{hollik,bigfit} of the standard model of the electroweak
interactions
\cite{sm} it is mainly  the fermion-gauge boson
couplings that have been tested and the gauge sector of the
standard model remains largely {\it terra incognita}.
A stringent test of the gauge structure of the standard model is provided by
the
tri-linear gauge vertices (TGV's); the $\gamma WW$ and $ZWW$ vertices. Within
the
standard model, these couplings are uniquely determined
by $SU(2)_L \times U(1)$ gauge invariance so that a precise measurement
of the vertex poses a severe test of the gauge structure of the theory.
If these couplings were observed to have different values than
their standard model values, it would indicate the need for physics beyond
the standard model.

A problem common to many processes used to study TGV's is that
they involve
both the $WW\gamma$ and $WWZ$ vertices making it difficult to disentangle
the contributions.  In
this paper we study the sensitivity of the process
$e^+e^-\to (Z, \gamma^*) \nu_l\bar\nu_l\to\mu^+\mu^-\nu_l\bar\nu_l$ to
anomalous
couplings in the $\gamma WW$ and $ZWW$ vertices.
This process offers the possibility of studying the $ZWW$ vertex
independently of the $\gamma WW$ vertex by imposing appropriate kinematic
cuts to select the invariant mass of the $\mu^+\mu^-$ pair. We start with
$\sqrt{s}=200$ GeV appropriate to LEP200 since this machine will be operational
in the
relatively near future \cite{lep200}. We then turn to the proposed JLC/NLC/CLIC
$e^+e^-$
colliders with possible center of mass energies of $\sqrt{s}=500$
GeV and  1 TeV\cite{nlc,JLC,NLC2,CLIC}. It is important to mention
that we do not include any beamsstrahlung radiation effects
in our calculation \cite{beam}. These effects are
very much machine dependant (beam intensity, bunch geometry, etc \dots)
and known to be
negligible at 200 GeV, and small at 500 GeV. However, although
they can be quite important at 1000 GeV,
there has been recent progress in strategies to minimize the
effects of beamstrahlung radiation.  Interestingly, such high
energy colliders offer the possibility of  studying the process $e\gamma\to
W^-\nu_e$ or $\gamma\gamma\to W^+W^-$
\cite{yehudai,choi91}.  Both of these processes have been
studied in detail  and appear very promising.

The outline of this paper is as follows:
In the next section we write down the effective Lagrangian
we will be studying and
the resulting Feynman rule to give our conventions.  We also discuss the
present constraints on TGV's and expected constraints from future
experiments.
In section III  we  examine in detail
the process we are interested in; $e^+e^-\to
\mu^+\mu^- \nu \bar{\nu}$ and describe the method of  calculation.
In section IV we present our results for the three
energy regimes that we investigated. We summarize
our conclusions in section V.

\section{THE $WWV$ EFFECTIVE VERTEX}
\label{sec:vertex}

A particularly useful means of probing for physics at high energy scales
is to use the language of effective Lagrangians\cite{bigfit,bagger}.  An
effective Lagrangian
parametrizes in as model-independent a way as possible the low-energy
implications of new physics at a much higher scale, $M$.  The effective
Lagrangian offers a common language so the sensitivity of various
experimental observables can be compared in a model-independent way.

There are several different effective Lagrangians in the literature used
to describe the trilinear gauge boson vertices (TGV's).  They differ in that
they
make different assumptions on the symmetries and particle content respected
by the effective Lagrangian.  In our analysis we use the most general
parametrization possible that respects Lorentz
invariance, electromagnetic gauge invariance,
and CP invariance\cite{hagiwara87,gaemers79,miscvertex}.
Because this general Lagrangian hides the $SU(2)\times U(1)$ symmetry
observed at present energies and obscures the expected size of it's
parameters, it has been the object of some
criticism in the literature\cite{derujula}.
It is, in fact, equivalent to the alternative $SU(2)\times U(1)$ invariant
nonlinearly realized Lagrangian written in the Unitary gauge upon
suitable field redefinitions \cite{equivalence} and in general one can
transform
the parameters of one effective Lagrangian to the parameters of another
\cite{falk}.
We choose to use the general Lagrangian
in our analysis since it has become the standard
parametrization used in phenomenology
and therefore makes the comparison of the sensitivity of different
observables to the TGV's  straightforward.

The most general $WWV$ vertex, satisfying Lorentz invariance, U(1)
gauge invariance and CP conservation allows four free independent
parameters, $\kappa_\gamma$, $\lambda_\gamma$, $\kappa_Z$ and $\lambda_Z$
when  the $W$ bosons couple to essentially massless fermions which
effectively results in $\partial_\mu W^\mu=0$ \cite{hagiwara87,gaemers79}.
We do not
consider CP violating operators in this paper as they  are tightly
constrained by measurement of the neutron
electron dipole moment which constrains the two CP violating parameters to
$|\tilde{\kappa}_\gamma|, |\tilde{\lambda}_\gamma|<
{\cal O} (10^{-4})$ \cite{cp}.
Therefore, the most general  Lorentz and CP invariant vertex compatible with
electromagnetic gauge invariance is commonly parametrized as
\cite{hagiwara87,gaemers79}:
\begin{equation}
{\cal L}_{WWV} =  - ig_V \left\{ { (W^\dagger_{\mu\nu}W^\mu V^\nu -
W^\dagger_\mu V_\nu W^{\mu\nu} )
+ \kappa_V W^\dagger_\mu W_\nu F^{\mu\nu}
- {{\lambda_V}\over{M_W^2}} W^\dagger_{\lambda\mu}W^\mu_\nu F^{\nu\lambda}
}\right\}
\end{equation}
where $V$ represents either the photon or the $Z^0$ and $W^\mu$ the $W^-$
fields. As usual, $W_{\mu\nu}=\partial_\mu W_\nu-\partial_\nu W_\mu$ and
$F_{\mu\nu}=\partial_\mu V_\nu-\partial_\nu V_\mu$ where $V$ is either the
photon or the Z boson,
$M_W$ is the $W$ boson mass, and $g_\gamma=e$ and $g_{Z^0}=e\cot\theta_w$.
Higher dimension operators would correspond to momentum dependence in the form
factors which we ignore.
At tree level the standard model requires $\kappa_V=1$ and $\lambda_V=0$.
Note that the presence of the W-boson mass factor in the
$\lambda_V$ term is {\it ad hoc} and one could argue that the scale $\Lambda$
of new physics would be more
appropriate. We will conform to the usual parametrization and will not address
this issue any further.

The resulting Feynman rule for
the $WWV$ vertex is given below with the notation
and conventions given in fig. 1.
\begin{eqnarray}
ig_V
\{ g_{\alpha\beta}[(1-\tilde{\lambda} \; k_- \cdot q ) {k_+}_\mu
-(1-\tilde{\lambda} \; k_+ \cdot q ) {k_-}_\mu ] & \nonumber \\
 - g_{\alpha\mu}[(1-\tilde{\lambda} \; k_- \cdot q ) {k_+}_\beta
-(\kappa-\tilde{\lambda} \; k_+ \cdot k_- ) q_\mu ] &
- g_{\beta\mu}[(\kappa-\tilde{\lambda} \; k_- \cdot k_+ ) q_\alpha
-(1- & \tilde{\lambda} \; k_+ \cdot q ) {k_-}_\alpha ] \nonumber \\
& +\tilde{\lambda} ({k_+}_\mu {k_-}_\alpha q_\beta
-{k_-}_\mu q_\alpha {k_+}_\beta ) \}
\end{eqnarray}
where $g_V=e$ for $V=\gamma$ and $e\cot\theta_w$ for $V=Z^0$ and
$\tilde{\lambda}=\lambda/M_W^2$.

In the static limit (all particles on mass-shell), the
parameters $\lambda_\gamma$ and $\kappa_\gamma$ are related to the anomalous
magnetic and electric quadrupole moments of the $W$ boson by:
$$ \mu_W={e\over 2 M_W}(1+\kappa_\gamma+\lambda_\gamma)
\qquad
Q_W = {-e\over 2M_W^2}(\kappa_\gamma-\lambda_\gamma) $$
with similar expressions for the weak moments (i.e. those that involve the Z
boson) At tree-level, the standard model requires $\kappa_V=1$ and
$\lambda_V=0$. Higher
order corrections to $\mu_W$ and $Q_W$ have been calculated in the past and
the results are in the 2\% range in the minimal standard model and in the
3\% range in
minimal supersymmetric extensions of the model\cite{vertexrc}.

Constraints can be obtained from precision measurements on the
$WW\gamma$ and $WWZ^0$ vertices via loop corrections
since deviations from their standard model values would
have resulted in discrepancies of observables
from their standard model predictions \cite{tgv,tgvmad,tgvothers}.
At present the
limits on TGV's obtained from a global analysis of precision measurements
are relatively weak; $|\delta \kappa_\gamma| \leq 0.12$,
$|\delta \kappa_Z| \leq 0.08$, $|\lambda_\gamma| \leq 0.07$, and
$|\lambda_Z| \leq 0.09$ at 95\% C.L. varying one parameter at a time\cite{tgv}.
In a simultaneous fit cancellations could lead to larger values and
in addition, because there are ambiguities in the
extraction of these bounds from loop calcuations due to ignorance of the
operators values at high energy and the scale of new physics, the
bounds obtained in this manner are at best order of magnitude estimates.

In contrast,  direct  measurements of  gauge boson couplings are
unambiguous. The only existing direct limits come
from the measurement of associated $\gamma W$
production by the UA2 experiment at the CERN $\bar{p}p$ collider
which obtained
$-3.5 < \kappa_\gamma < 5.9 $ and $ -3.6 < \lambda_\gamma < 3.5$
at 95\% C.L. \cite{ua2}. The limits obtained from the Tevatron are unsettled
at present, with two theory analysis finding significantly different
limits \cite{tevatron1,tevatron2}.  The most optimistic limits from the
Tevatron are $|\delta\kappa_\gamma|\simeq 3$
and $|\delta\lambda_\gamma| \simeq 1.2$ at 68\% C.L. \cite{tevatron2}
The sensitivities expected at
an upgraded Tevatron with $L=100\hbox{pb}^{-1}$ are
$|\delta\kappa_\gamma|\simeq 1.4 $  and
$|\delta\lambda_\gamma|\simeq 0.47$ at 90\% C.L.\cite{tevatron2}.
In the
near future, HERA will be able to constrain the $\gamma WW$ vertex through
single W production\cite{heraw1,heraw2,heraw3}
and high $p_T$ photons\cite{herag}.
Statistics will be the main
limiting factor and a precision of $\pm 0.5$ or so is expected\cite{heraw1}.

Putting tight constraints on the trilinear
gauge boson couplings by studying $W$ pair production
is one of primary motivations for the LEP200 upgrade
\cite{hagiwara87,lep200,kane89}. A precision of 30-40\% is expected
from a direct measurement of the cross-section. If one can reconstruct the
W-bosons, their angular distribution offers a more sensitive probe and could
lead to a bound of 25\% or so. Another possibility it to study single $W$
production in $e-\gamma$ collisions \cite{couture,yehudai,choi91}.
The process $\gamma \gamma\to W^+W^-$ through heavy ion collisions
also offers interesting possibilities\cite{heavyion}.
However, one has to deal with an enormous background in the case of head-on
collisions or a greatly reduced hard-photon rate for glancing collisions.
In the longer term
the LHC offers good possibilities. Baur and Zeppenfeld\cite{ssc}
have shown that a measurement of $\vert\delta\kappa_\gamma\vert\sim 0.2-0.5$ at
99.9\% cl. or better is possible, assuming
an integrated luminosity of $10^4$ pb$^{-1}$.

In the far future there is growing interest in the physics that can be done
at high energy $e^+e^-$ colliders with $\sqrt{s}=500$ GeV or $\sqrt{s}=1$ TeV,
referred to as  the Next Linear
Collider (NLC), the Japan Linear Collider (JLC)  or the CERN Linear Collider
(CLIC) \cite{nlc,JLC,NLC2,CLIC}.
Various options are being studied including $e\gamma$ collisions
where the energetic photons are obtained either  by backscattering a laser on
one of the incident leptons or by beamstruhlung photons.
Measurements at these colliders are very sensitive to anomalous
couplings with $e\gamma$ collisions
putting some of the more stringent bounds on anomalous $WW\gamma$
couplings \cite{yehudai,choi91}.

\section{CALCULATIONS AND RESULTS}
\label{sec:process}

The process $e^+e^-\to(\gamma^*Z)\nu_l\bar\nu_l\to\mu^+\mu^-\nu_l\bar\nu_l$
has several advantages; it is a t-channel process and  does not
decrease as quickly as s-channel processes as the c.m. energy increases. More
importantly, it offers the possibility  of isolating the $ZWW$ vertex
from the $\gamma WW$ vertex by imposing appropriate cuts on the
invariant mass of the $\mu^+\mu^-$ pairs. One drawback is that a total of 28
Feynman diagrams
contribute to the process  and one has to add incoherently the three families
of neutrinos.
Although only the 2 diagrams shown in fig.2 (a) and 2(b) contribute to
the vertex we wish to study, in order to
properly take into account the non resonant backgrounds and maintain gauge
invariance, at least in the standard model limit,
we must  include all 28 diagrams. We leave $\kappa_V$ and
$\lambda_V$ as free parameters.

To evaluate the cross-sections and different distributions, we used the CALKUL
helicity amplitude technique \cite{calkul} to obtain expressions for the
matrix elements and
performed the phase space integration using  Monte Carlo
techniques \cite{monte}.
The expressions for the helicity amplitudes are lengthy and unilluminating so
we do not include them here.  The interested reader can obtain them directly
from the authors.
To obtain numerical results we used the values $\alpha=1/128$,
$\sin^2\theta=0.23$, $M_Z=91.187$ GeV, $\Gamma_Z=2.5$ GeV, $M_W=80.2$ GeV, and
$\Gamma_W=2.1$ GeV.

The signal we are studying
is an energetic $\mu^+\mu^-$ pair plus missing transverse momentum
due to the neutrinos coming from the original beams. In order to eliminate
potential background from $e^+e^-\to e^+e^-\mu^+\mu^-$ via two photons, where
the $e^+$ and $e^-$ escape down the beam pipe, we require missing
${\sla{p_T}}\geq$ 10 GeV.
We also require $10\leq E_{\mu^\pm}\leq\sqrt{s}/2-10$ GeV, to avoid
2 body s-channel processes and box diagrams. Note that these kinematic
cuts overlap.  In order to take into account finite detector acceptance, we
require that the $\mu^+$ and $\mu^-$ be at least 10 degrees away from the beam
line.
Our conclusions are not very sensitive to the exact values of these cuts.

In
fig. 3, we show the cross-section for $e^+e^-\to \nu_l\bar{\nu}_l\mu^+\mu^-$ as
a
function of $\sqrt{s}$ with the cuts described above and consider the
effects of different cuts on
the invariant mass of the $\mu^+\mu^-$ pairs. We note that,
as expected of typical t-channel behavior,  the cross-section
does not go down with energy as does the QED point cross-section.
The cross-section for no cuts on $M_{\mu^+\mu^-}$ (solid line) is considerable
but it is dominated by low invariant mass events due to the photon pole which
appears in many of the Feynman diagrams including photon bremsstrahlung and the
diagram of interest in  fig 2(a).   For the most part the low invariant mass
contributions are an unwanted background which obscures the physics we are
interested in.
Imposing a cut of $M_{\mu^+\mu^-}>25$ GeV eliminates this pole and
reduces the cross-section substantially (long-dashed line). Finally, if we
impose the cut that $M_{\mu^+\mu^-}$ lies within 5 GeV of the Z pole
(dotted line) we can separate the effects of the ZWW vertex from the
$\gamma WW$ vertex since the photon contribution is now smaller by a factor
$(\Gamma_Z/M_Z)^2$. This last curve is in fact the superposition of two
diagrams:
the s-channel  process of fig. (2b) which  rises sharply above 200 GeV and
falls quickly at
$\sqrt{s}\geq 300$ GeV and  the W fusion diagrams
that fall quickly below 200 GeV and rise up to 1000 GeV. The
cut $M_{\mu^+\mu^-}\geq 25$ GeV is an intermediate state between this
extreme
and the photon-dominated case where the photon bremstrahlung diagrams dominate
for the entire energy range.

It is clear from this figure that this process is
hopeless for LEP-100; without any cut on $M_{\mu^+\mu^-}$ the cross-section is
substantial, but rather insensitive to variations on $\kappa_V$ and
$\lambda_V$. Imposing a cut on the invariant mass increases the
sensitivity to anomalous couplings but reduces the cross-section to
an unmeasurable level.



\newpage
\begin{center}
A. $\sqrt{s}=200$ GeV
\end{center}

\vskip 0.2cm

For $\sqrt{s}=200$ GeV we use the kinematic cuts $5 < E_{\mu^\pm}<95$ GeV
and $\sla{p_T} >5$ GeV.  The cross section with these cuts and in addition,
cuts on $M_{\mu^+\mu^-}$ of no cut,
$M_{\mu^+\mu^-}> 10$ GeV, and 86 GeV $<M_{\mu^+\mu^-}<$ 96 GeV are
1.9 pb, 0.23 pb, and 0.035 pb respectively.
The latter cut would isolate the effects of the $WWZ$ vertex from that
of the $WW\gamma$ vertex.  Unfortunately, for the
parameters of LEP-200 ($\sqrt{s}=200 GeV$, and an integrated
luminosity of 250 $pb^{-1}$) the number of events remaining after these
cuts  is not statistically useful.

To maximize the sensitivity to anomalous
TGV couplings we examined numerous kinematic
distributions.  The two which best separated the uninteresting photon
bremstrahlung
contribution from signals for anomalous couplings are $d\sigma/dM_{\mu^+\mu^-}$
and $d\sigma/d\cos\theta_{\mu^+\mu^-}$ which are shown in fig. 4 for
several values of $\kappa_\gamma$, $\lambda_\gamma$, $\kappa_Z$ an
$\lambda_Z$.
It is clear from fig. (4a) that sensitivity to anomalous couplings is due to
interference between the photon and $Z^0$ propagators.  We therefore
examined the effects of removing the contribution of the $Z^0$ pole;
$10 <M_{\mu^+\mu^-} < 88$ GeV.
Although this increased the ``signal to background'' it also
reduced the cross section. Once realistic efficiencies are considered
we do not feel that enough events would be left  to improve the measurement
of the TGV's.  We included the cuts of $M_{\mu^+\mu^-}> 10$ GeV  and
$\cos\theta_{\mu^+\mu^-}<0.95$ in our subsequent calculations used to
determine the sensitivity of  the measurements to anomalous couplings.
We note that these two cuts correspond to the same region in phase space
and therefore overlap.  In addition, deviations from the standard model
show up in various distributions such as $d\sigma/d\cos\theta_{\mu^+\mu^-}$
and we could bin these distributions to perform a $\chi^2$ analysis.
In practice, however, the cross section is too small at $\sqrt{s}=200$ GeV
to improve the sensitivity.

Since there are four free parameters, the parameter space is four
dimensional which can be projected onto six 2-dimensional planes. We
performed an extensive search in the parameter space and found that to a good
approximation the
largest ellipse in any given plane is reached when the other two parameters
are kept at their standard model values. The small exception is the relative
insensitivity
to variations in $\kappa_\gamma$ which results  in a very small enlargement of
the boundary for non-standard model values along  the $\kappa_\gamma$ axis.
The 95\% C.L. for integrated luminosities of 250~pb$^{-1}$ and
500~pb$^{-1}$ for the two planes, $\kappa_Z$
versus $\lambda_Z$  and $\kappa_\gamma$ versus $\lambda_\gamma$ are shown in
fig. 5.
These bounds represent the regions of the parameter space that can be ruled
out as inconsistent with the standard model for a measurement of the
standard model values for the given integrated luminosity.
If we vary one parameter at a time and hold the rest at their standard model
values we obtain the limits, based on the statistical error obtained from an
integrated luminosity of 500 pb$^{-1}$, $\delta\kappa_\gamma= ^{+1.9}_{-1.0}$,
$\delta\lambda_\gamma=^{+0.9}_{-1.3}$,
$\delta\kappa_Z =\pm 1.0$,  $\delta\lambda_Z =\pm 0.8$ at 95\% C.L..
In this analysis we did not assume any constraints on the parameters.
Imposing a custodial $SU(2)$ symmetry
gives the relation $\lambda_Z\simeq\lambda_\gamma$ \cite{kuroda}.
If we also take $\kappa_Z=\kappa_\gamma$ we obtain the sensitivities
$\delta\lambda=^{+0.6}_{-0.7}$ and $\delta\kappa = ^{+1.0}_{-0.6}$
at 95\% C.L. which is not so different
from the unconstrained result for the $Z$ parameters but
significantly tighter than the unconstrained results we obtained for
$\kappa_\gamma$ and $\lambda_\gamma$.
Since the contour axes are almost aligned with the parameter axes
two planes contain most, if not all the information about
the limits on the 4 parameters.

At 200 GeV we find that the constraints that can be achieved using the
process $e^+e^-\to \mu^+\mu^-\nu\bar{\nu}$
cannot compete with the constraints obtained from W-pair production.
Therefore, at best this process would
provide a consistency check for other measurements.


\vskip 0.3cm
\begin{center}
B. $\sqrt{s}=500$ GeV
\end{center}

\vskip 0.2cm

We next turn to an ``NLC'' type $e^+e^-$ collider with $\sqrt{s}=500$
GeV.  We consider integrated luminosities of 10 and 50 $fb^{-1}$ and
use the kinematic cuts 10~GeV $<E_{\mu^\pm} < 240$~GeV
and $\sla{p_T} > 10$ GeV.
With these cuts we obtain the invariant mass distribution,
$d\sigma/dM_{\mu^+\mu^-}$ shown in fig. 6.

The increased cross section and expected high luminosity at the NLC
leads to a significantly larger number of events making it possible
to study the reaction at
the $Z^0$ pole, significantly reducing the contributions of
the $\gamma WW$ vertex to the process.  Although results have been presented
previously for $|M_{\mu^+\mu^-}-M_Z|<5$~GeV \cite{eptoz1} the lumininosity
that the NLC is expected to achieve is significantly higher than what was
used in the earlier analysis.  We have therefore revised the analysis taking
into account the higher luminosity and emphasizing its implications
before proceeding to the results off the $Z^0$ resonance.
In particular we will see that systematic errors will play an increasingly
important role in precision measurments.

For the cut $|M_{\mu^+\mu^-}-M_Z|<5$~GeV we verified that the
cross section and distributions are insensitive to variations in
$\kappa_\gamma$ and $\lambda_\gamma$.
We considered the effects of varying $\kappa_Z$ and
$\lambda_Z$ on the cross section $\sigma (e^+e^- \to \mu^+\mu^-
\nu_l\bar{\nu}_l)$ and found that varying one parameter at a time
we obtained a sensitivity (in the sense of consistency with the standard
model)  of $\delta\kappa_Z=\pm 0.1$
($\lambda_Z=0$) and $\delta\lambda_Z={+0.2\atop -0.5}$ ($\kappa_Z =1$)
at 95\% C.L. based on 20 fb$^{-1}$ integrated luminosity.
However,  when we let both parameters vary at the same time we find
that regions in the parameter space very far from the standard model
give cross sections consistent with the standard model value \cite{eptoz1}.
To eliminate the ambiguities we examined a number of kinematic distributions.
The most sensitive are
the angular distribution of the muons with respect to each other
($\theta_{\mu^+\mu^-}$) and the
transverse momentum of the reconstructed $Z^0$ boson (${p_T}_Z$)
which we show in fig. 7 for several values of $\kappa_Z$ and $\lambda_Z$.
We performed a $\chi^2$ analysis based on the angular distribution
using the bins;
$-1.0<\theta_{\mu^+\mu^-}<-0.5$,
$-0.5<\theta_{\mu^+\mu^-}-0.1$, and $-0.1<\theta_{\mu^+\mu^-}< 0.75$ and
another one based on the ${p_T}_Z$ distribution with the bins;
${p_T}_Z <80$ GeV, 80 GeV $<{p_T}_Z <120$ GeV, and 120 GeV $<{p_T}_Z <240$
GeV.  The 68\%, 90\%, and 95\% C.L. bounds using
$d\sigma/d\cos\theta_{\mu^+\mu^-}$ and $d\sigma/d{p_T}_Z$
based on 10~fb$^{-1}$ are shown
in fig. (8a) and (8b) respectively. This additional information substantially
restricts the allowed region in parameter space that is consistent with
the standard model with $\delta \kappa_Z=\pm 0.2$ and
$\delta\lambda=^{+0.3}_{-0.4}$ at 95\% C.L.  .

These results were based on 10~fb$^{-1}$ of integrated luminosity.  It is
expected that the luminosity is likely to be higher than this which would
improve the measurement capabilities.  On the other hand we have
neglected systematic errors in our analysis.  Monte Carlo studies of SLD
type detectors give very crude estimates of systematic errors of 5\% for
cross section measurements \cite{barklow}.
In fig. 9 we show the 95\% C.L. assuming
10~fb$^{-1}$ and 50~fb$^{-1}$ with and without a 5\% systematic measurement
error.  We find that although improving the statistical
error improves the sensitivity the systematic error tends to be more
important.  In other words, one gains more by reducing the systematic error
than by significantly increasing the luminosity.

Although these results are
no improvement over the expected LEP200 measurements  based on $W^+W^-$
pair production, they offer a means of
measuring the $WWZ$ vertex independently of the $WW\gamma$ vertex.

Whereas isolating the  $Z^0$ pole offers a means of studying
the $WWZ$ vertex independently of
the $WW\gamma$ vertex, there is a severe penalty in terms of reduced
cross section and sensitivity to anomalous coupling.  We therefore examine
less restrictive cuts on the $\mu^+ \mu^-$ invariant mass
which restores the $WW\gamma$ vertex.
We searched for the range of $M_{\mu^+\mu^-}$ which exhibited the largest
sensitivity to anomalous couplings and found it to be
$110\leq M_{\mu^+\mu^-}\leq 400$ GeV.
As before, with four
independant parameters, an extensive search in the parameter space
showed that,
to a good approximation,  the largest (weakest) confidence limit bounds in any
two
parameters are reached when the
other two parameters are kept at their standard model values. The contours for
$\kappa_\gamma$ vs $\lambda_\gamma$ and $\kappa_Z$ vs $\lambda_Z$ are
shown in fig. (10a) and (10b) respectively, based on the statistics from
10~fb$^{-1}$ integrated luminosity.
These bounds were improved slightly when we relaxed the acceptance cuts
around the beam axis to $|\cos\theta|<0.8$ from $\cos\theta>10^o$.
We examined other kinematic distributions
but found that for the parameter range allowed by fig. 10,
the distributions
are very similar and did not offer a significant improvement of the bounds
obtained from the cross section measurement.  (Imposing the kinematic cuts
$E_Z<300$ GeV gave a very slight improvement and ${p_T}_Z>100$ GeV distorted
the ellipses to give a slight improvement on $\delta \lambda_Z$)

Finally, we binned the
$\mu^+\mu^-$ invariant mass distribution into the four bins;
$25<M_{\mu^+\mu^-}<86$, $86<M_{\mu^+\mu^-}<96$,
$96<M_{\mu^+\mu^-}<110$,
and $110<M_{\mu^+\mu^-}<400$ and performed a $\chi^2$ analysis,
varying one parameter at a time.  The resulting  95\%
C.L. bounds based on L=10~fb$^{-1}$ and L=50~fb$^{-1}$ with and without
systematic errors are summarized in Table I along with our other results
for $\sqrt{s}=500$~GeV.
As before, one should be cautious about possible ambiguities when
intepreting the results  obtained by varying only one parameter at a time.
Only the
results for L=50~fb$^{-1}$ approach the sensitivity required to observe loop
contributions to the TGV's.  However, when systematic errors are included it
is unlikely that these measurements will reveal non-standard model physics
through radiative corrections to the TGV's.  In addition the systematic
errors are the limiting factor in the sensitivities, not the statistical
errors.


\vskip 0.3cm

\begin{center}
C. $\sqrt{s}=1$ TeV
\end{center}

\vskip 0.2cm

The final case we consider is a 1 TeV $e^+e^-$ collider.
Although beamsstrahlung
effects are known to be important in this energy regime
\cite{beam}, we will neglect them
since they depend on many machine dependent factors making it difficult to
estimate at this time.  In any case, much progress has been made in
understanding how to eliminate beamstrahlung so it may not be as important a
factor as originally feared.  This should be kept
in mind when assessing our results.
In what follows
we use the kinematic cuts of $\sla{p_T}$ of 10 GeV, $\theta_{\mu^\pm}>10^o$,
and $10<E_{\mu^\pm}<490$~GeV.

As before, we imposed a cut on $M_{\mu^+\mu^-}$ to isolate the $Z^0WW$ vertex.
The standard model cross-section is 0.52 pb.
We verified explicitly that varying
$\kappa_\gamma$ and $\lambda_\gamma$ by 10\% changed the total cross-section by
2 part in 10,000 or less as expected. Varying $\kappa_Z$ and
$\lambda_Z$ by 10\% changed the total cross-section by 2\% or less.
This small variation does not translate into particularly significant bounds
on the TGV's.
As before we performed a $\chi^2$ analysis based
on  four equal bins for  $\theta_{\mu^+\mu^-}$ and the four bins
${p_T}_Z <125$ GeV, 125 GeV $<{p_T}_Z <250$ GeV,
250 GeV $<{p_T}_Z <375$ GeV, and ${p_T}_Z >375$ GeV.
The 68\%, 90\%, and 95\% C.L. bounds using
$d\sigma/d\cos\theta_{\mu^+\mu^-}$ and $d\sigma/d{p_T}_Z$
based on 50~fb$^{-1}$ are shown in fig. (11a) and (11b) respectively.
Although the TGV's can be constrained to less than about 0.07 with
50~fb$^{-1}$ and less than about 0.04 with 200~fb$^{-1}$, when a 5\%
systematic error is included these bounds weaken to about 0.12.

Given these relatively weak bounds we concentrated on large
invariant mass where there can be interference between the two vertices
so that although the
cross-section is reduced, the sensitivity to anomalous coupling is greatly
increased.
As seen on fig. 12, the greatest sensitivity to anomalous coupling occurs in
the mass range $200\leq M_{\mu^+\mu^-}\leq 900$ GeV although the upper
bound could be pushed to 1 TeV without changing our results. Note that the
values used here ($\delta\kappa=\delta\lambda=\pm 0.2$) are extreme
and are used simply to illustrate our point.
The resulting standard model cross-section is 0.024~pb.
We show on fig.  13 two of the six planes of the four
dimensional parameter space. Over this small range in parameter space, the
shapes of the different distributions are very similar and one would not gain
much (if anything at all) by considering them over a total cross-section
measurement. The 95\% C.L. bounds that we obtain assuming 50~fb$^{-1}$ and
no systematic errors, varying one parameter at a time are:
$\delta\kappa_\gamma ={+0.04\atop -0.01}$,
$\delta\kappa_Z =\pm 0.01$,
$\delta \lambda_\gamma ={+0.03\atop -0.01}$,
and $\delta \lambda_Z {+0.02\atop -0.01}$.
For comparison, the bounds obtained by assuming
$\kappa_Z=\kappa_\gamma$ and $\lambda_Z=\lambda_\gamma$
imposed by custodial SU(2) symmetry\footnote{Strictly speaking
custodial SU(2) imposes $\lambda_Z=\lambda_\gamma$ and
$\kappa_Z=\kappa_\gamma=1$ \cite{kuroda}.}
are slightly stronger with
$\delta\kappa= {+0.03\atop -0.006}$
and$\delta\lambda ={+0.008\atop -0.006}$. Again, we
emphasize that one must be cautious in interpreting bounds obtained by
varying one parameter at a time.
The bounds that can be obtained
with the various kinematic cuts, luminosities, with and without
systematic errors are summarized in Table II.
It is clear that the greatly improved sensitivity to anomalous coupling more
than compensates for the reduced cross-section.

\section{CONCLUSIONS}

We have shown that the process $e^+e^-\to \mu^+ \mu^- \nu \bar{\nu}$
can be useful in disentangling the
different contributions to the $\gamma W^+W^-$ and $Z^0W^+W^-$ vertices.
We included all processes that lead to this four-fermion final state.
Appropriate
cuts on the invariant mass of the muon pairs offer the possibility of
measuring the $Z^0W^+W^-$ vertex by itself albeit with reduced precision.
This is due to both reduced statistics but also due to reduced sensitivity
of the process to anomalous couplings.  With the cut
$|M_{\mu^+\mu^-}-M_Z|<5$~GeV the cross section is dominated by the one
diagram and gauge vertex we are interested in so that
there are no sensitive gauge cancellations in the process.
On the other hand, off the $Z^0$ resonance the anomalous
couplings can interfere with other diagrams resulting in greater sensitivity
to anomalous couplings.

The process turned out to be hopeless at LEP-100 because of the
low cross section. At LEP-200 it can lead to, at best,
a consistency check of bounds extracted from $W$-pair production.
At higher energy
$e^+e^-$ colliders, it can lead to very stringent bounds precise enough to
test the TGV's at the level of radiative corrections. The bounds were
obtained using many different measurements such as the angular distributions
of the outgoing muons, the transverse momentum distribution of the
reconstructed $Z^0$ boson, and the integrated cross section for the process
{\it off} the $Z^0$ resonance.

Studying the four-fermion final state and including all diagrams which
contributes to the final state leads to a much richer phenomenology than
would be obtained by simply studying final state gauge bosons.  Thus, the
entire process, with all the contributing diagrams, should be studied when
examining the physics potential of a specific reaction.
We used the
high luminosities planned for at the high energy $e^+e^-$ colliders to
estimate statistical errors.  However, when we included reasonable estimates
of systematic errors we found that the limiting factor in high precision
measurements will likely be systematic errors not statistical errors.
The challenge will therefore
be to reduce the systematic errors and one should be very
careful with respect to the conclusions one makes by only considering
statistical errors.

\acknowledgments

This research was supported in part by the Natural Sciences and
Engineering Research Council of Canada and Les Fonds FCAR du
Quebec. The authors gratefully acknowledge the contributions of
Randy Lewis in the early stages of this work.
G.C. wants to thank the Superconducting
Supercollider Laboratory for the use of their computing facilities.  S.G.
thanks Tim Barklow and Dean Karlen for helpful conversations and Drew
Peterson for assistance in preparing the figures.

\newpage
\centerline{The Helicity Amplitudes}

In this appendix\footnote{Note: This
appendix does not appear in the version submitted to the
journal.}, we outline the use of the CALKUL
spinor technique.
We limit our discussion to
massless fermions which apply to our problem.
The propagators for the fermions and gauge bosons have the same form as in
the trace technique so  we do not discuss them here.

The spinor technique results in reducing strings of spinors
and gamma matrices to sandwiches of spinors which can be evaluated easily.
In doing so, one makes extensive use of the {\it left} and {\it right}
projection operators defined by
$ \omega_{\pm} = {1\over2}(1{\pm}\gamma_5 ) $.
One defines two four-vectors, $k_0^{\mu}$ and $k_1^{\mu}$, which obey the
following relations:
$$k_0\cdot k_0 = 0, \phantom{ssss}k_1\cdot k_1 = -1,\phantom{ssss}k_0\cdot
k_1 = 0 $$
and the basic spinors as:
$$ u_{-}(k_0){\bar u}_{-}(k_0) = {\omega}_{-}\sla{k_0} $$
and
$$ u_{+}(k_0) = \sla{k_1} u_{-}(k_0) .$$
Note that in the massless limit, one can use $u$ and ${\bar u}$ to describe
both particles and antiparticles, with the spin sum
$\sum_{\lambda} u_\lambda (p) {\bar u}_\lambda (p) = \sla{p} $.
These two spinors are the building blocks for any spinor of lightlike
momentum p :
$$ u_{\lambda}(p)= { { \sla{p}u_{-\lambda}(k_0)} \over
{\sqrt{2\;p\cdot k_0} } }$$
Two identities are essential for the reduction of the strings; the spin sum
given above and  the Chisholm identity:
$$ {\bar u}_\lambda (p_1){\gamma}^{\mu} u_\lambda (p_2) {\gamma}_{\mu}
   \equiv 2 u_\lambda (p_2) {\bar u}_\lambda (p_1) +
	  2 u_{-\lambda}(p_1) {\bar u}_{-\lambda}(p_2)$$
where ${\lambda}$ is ${\pm 1}$ and represents the helicity state.
These two identities allow one to reduce strings of spinors and gamma
matrices to sandwiches of spinors.
Only two of the four possible sandwiches are
non-zero:
$$ s(p_1,p_2)\equiv {\bar u}_+(p_1) u_-(p_2) = -s(p_2,p_1) $$
and
$$ t(p_1,p_2)\equiv {\bar u}_-(p_1) u_+(p_2) = s(p_2,p_1)^{*}. $$
Once the amplitude has been reduced to a series of factors of $s( p_i ,p_j )$
and $t( p_k,p_l )$, the expressions can be
evaluated
by computer. A judicious choice of the four-vectors $k_0^{\mu}$ and
$k_1^{\mu}$ simplifies the evaluation of the $s$ and $t$ terms. For our
calculation, we use;
$$ p_i^{\mu} = ( p_i^0,p_i^x,p_i^y,p_i^z ) $$
$$ k_0^{\mu} = ( 1,1,0,0 ) $$
$$ k_1^{\mu} = ( 0,0,1,0 ) $$
to obtain
$$s(p_1,p_2) = (p_1^y + i p_1^z )
{\sqrt{ p_2^0 - p_2^x}\over{\sqrt{ p_1^0 - p_1^x}}} -
( p_2^y + i p_2^z )
{\sqrt{ p_1^0 - p_1^x}\over{\sqrt{ p_2^0 - p_2^x}}} $$

These forms are ideally suited for programming. When dealing with several
diagrams, one simply evaluates the amplitudes of each diagram
as complex numbers and squares the sum of the  amplitudes in order to
obtain the $\vert amplitude \vert ^2$.

Using this technique and the following definitions
$$
e^-(p_1)+e^+(p_2)\to \mu^-(p_3)+\mu^+(p_4)+\nu_l(p_5)+\bar\nu_l(p_6)
$$
we obtain for the helicity amplitudes,

\begin{eqnarray*}
M_{LL}^{a} &=& {-2ig^2g_V^2}
\; C_L^{Ve}
\; D_W(p_1-p_5)
\; D_W(p_2-p_6)
\; D_V(p_3+p_4) \\
& & \qquad \{ t(5,2)\;s(6,1)*
[(1.0+\tilde{\lambda}_V \; p_{W^+}\cdot
p_{Z})*(t(3,1)\;s(1,4)-t(3,5)\;s(5,4))\\
& & \qquad
-(1.0+\tilde{\lambda}_V \; p_{W^-}\cdot p_{Z})*
	(t(3,2)\;s(2,4)-t(3,6)\;s(6,4))]\\
& & \qquad + t(5,3)\;s(4,1)*
[(\kappa_V+\tilde{\lambda}_V \; p_{W^-}\cdot p_{W^+})*
(-t(2,4)\;s(4,6)-t(2,3)\;s(3,6))\\
& & \qquad
-(1.0+\tilde{\lambda} p_{W^+}\cdot p_{Z})*(t(2,1)\;s(1,6)-t(2,5)\;s(5,6))]\\
& & \qquad + t(2,3)\;s(4,6)*
[(1.0+\tilde{\lambda}_V p_{W^-}\cdot p_{Z})*(t(5,2)\;s(2,1)-t(5,6)\;s(6,1))\\
& & \qquad
-(\kappa_V+\tilde{\lambda}\;p_{W^+}\cdot p_{W^-})*
(-t(5,4)\;s(4,1)-t(5,3)\;s(3,1))]\\
& & \qquad - {1\over 2}\tilde{\lambda}_V*
 [t(5,2)\;s(2,1)-t(5,6)\;s(6,1)]*[-t(2,4)\;s(4,6)-t(2,3)\;s(3,6)]\\
& & \qquad
 *[t(3,1)\;s(1,4)-t(3,5)\;s(5,4)]\\
& & \qquad + {1\over 2}\tilde{\lambda}_V
 *[-t(5,4)\;s(4,1)-t(5,3)\;s(3,1)]*[t(2,1)\;s(1,6)-t(2,5)\;s(5,6)]\\
& & \qquad
 *[t(3,2)\;s(2,4)-t(3,6)\;s(6,4)]\}\\
M_{LR}^{a} &=& {-2ig^2g_V^2}
\; C_R^{Ve} \; D_W(p_1-p_5) \; D_W(p_2-p_6) \; D_V(p_3+p_4) \\
& & \qquad \{t(5,2)\;s(6,1)
[(1.0+\tilde{\lambda}_V p_{W^+}\cdot p_{Z})*(s(3,1)\;t(1,4)-s(3,5)\;t(5,4))\\
& & \qquad
-(1.0+\tilde{\lambda}_V p_{W^-}\cdot p_{Z})*(s(3,2)\;t(2,4)-s(3,6)\;t(6,4))]\\
& & \qquad  +t(5,4)\;s(3,1) *
[(\kappa_V+\tilde{\lambda}_V p_{W^-}\cdot p_{W^+})
*(-t(2,4)\;s(4,6)-t(2,3)\;s(3,6))\\
& & \qquad
-(1.0+\tilde{\lambda}_V  p_{W^+}\cdot p_{Z})*(t(2,1)\;s(1,6)-t(2,5)\;s(5,6))]\\
& & \qquad + t(2,4)\;s(3,6) * [(1.0+\tilde{\lambda}_V p_{W^-}\cdot p_{Z})
*(t(5,2)\;s(2,1)-t(5,6)\;s(6,1))\\
& & \qquad
+(\kappa_V+\tilde{\lambda}_V p_{W^-}\cdot p_{W+})*
	(t(5,4)\;s(4,1)+t(5,3)\;s(3,1))]\\
& & \qquad - {1\over 2}\tilde{\lambda}_V
 *[t(5,2)\;s(2,1)-t(5,6)\;s(6,1)]*[-t(2,4)\;s(4,6)-t(2,3)\;s(3,6)]\\
& & \qquad
 *[s(3,1)\;t(1,4)-s(3,5)\;t(5,4)] \\
& & \qquad + {1\over 2}\tilde{\lambda}_V *
 [-t(5,4)\;s(4,1)-t(5,3)\;s(3,1)]*[t(2,1)\;s(1,6)-t(2,5)\;s(5,6)]\\
& & \qquad
  *[s(3,2)\;t(2,4)-s(3,6)\;t(6,4)]\} \\
M_{LL}^{b} & = & {-ig^2g_V^2}
\; C_L^{Ve}
\; D_W(p_3+p_6)
\; D_W(p_4+p_5)
\; D_V(p_1+p_2) \\
& & \qquad \{
 t(3,5)\;s(4,6)*[
(1.0+\tilde{\lambda}_V p_{W^+}\cdot p_{Z})*(-t(2,3)\;s(3,1)-t(2,6)\;s(6,1))\\
& & \qquad
-(1.0+\tilde{\lambda}_V p_{W^-}\cdot p_{Z})*(-t(2,4)\;s(4,1)-t(2,5)\;s(5,1))]\\
& & \qquad +t(2,3)\;s(6,1) *
[(\kappa_V+\tilde{\lambda}_V p_{W^-}\cdot p_{W+})*
	(t(5,2)\;s(2,4)+t(5,1)\;s(1,4))\\
& & \qquad
-(1.0+\tilde{\lambda}_V p_{W^+}\cdot p_{Z})*(-t(5,3)\;s(3,4)-t(5,6)\;s(6,4))]\\
& & \qquad + t(5,2)\;s(1,4) *
[(1.0+\tilde{\lambda}_V p_{W^-}\cdot p_{Z})*(-t(3,4)\;s(4,6)-t(3,5)\;s(5,6))\\
& & \qquad
-(\kappa_V+\tilde{\lambda}_V p_{W^-}\cdot p_{W+})*
	(t(3,2)\;s(2,6)+t(3,1)\;s(1,6))]\\
& & \qquad - {1\over 2}\tilde{\lambda}_V
 *[t(2,3)\;s(3,1)+t(2,6)\;s(6,1)]*[t(3,4)\;s(4,6)+t(3,5)\;s(5,6)]\\
& & \qquad
 *[t(5,2)\;s(2,4)+t(5,1)\;s(1,4)] \\
& & \qquad + {1\over 2}\tilde{\lambda}_V
 *[t(2,4)\;s(4,1)+t(2,5)\;s(5,1)]*[t(3,2)\;s(2,6)+t(3,1)\;s(1,6)]\\
& & \qquad
 *[t(5,3)\;s(3,4)+t(5,6)\;s(6,4)]\} \\
M_{RL}^{b} &=& {-ig^2g_V^2}
\; C_R^{Ve}
\; D_W(p_3+p_6)
\; D_W(p_4+p_5)
\; D_V(p_1+p_2) \\
& & \qquad \{
 t(3,5)\;s(4,6) *
[(1.0+\tilde{\lambda}_V p_{W^+}\cdot p_{Z})*(-s(2,3)\;t(3,1)-s(2,6)\;t(6,1))\\
& & \qquad
-(1.0+\tilde{\lambda}_V p_{W^-}\cdot p_{Z})*(-s(2,4)\;t(4,1)-s(2,5)\;t(5,1))]\\
& & \qquad + t(3,1)\;s(2,6) *
[(\kappa_V+\tilde{\lambda}_V p_{W^-}\cdot p_{W+})*
	(t(5,2)s\;(2,4)+t(5,1)\;s(1,4))\\
& & \qquad
-(1.0+\tilde{\lambda }_V p_{W^+}\cdot
p_{Z})*(-t(5,3)\;s(3,4)-t(5,6)\;s(6,4))]\\
& & \qquad + t(5,1)\;s(2,4) *
[(1.0+\tilde{\lambda}_V p_{W^-}\cdot p_{Z})*(-t(3,4)\;s(4,6)-t(3,5)\;s(5,6))\\
& & \qquad
-(\kappa_V+\tilde{\lambda}_V p_{W^-}\cdot p_{W+})*
	(t(3,2)\;s(2,6)+t(3,1)\;s(1,6))]\\
& & \qquad -{1\over 2} \tilde{\lambda}_V
*[s(2,3)\;t(3,1)+s(2,6)\;t(6,1)]*[t(3,4)\;s(4,6)+t(3,5)\;s(5,6)]\\
& & \qquad
 *[t(5,2)\;s(2,4)+t(5,1)\;s(1,4)]\\
& & \qquad +{1\over 2} \tilde{\lambda}_V
[s(2,4)\;t(4,1)+s(2,5)\;t(5,1)]*[t(3,2)\;s(2,6)+t(3,1)\;s(1,6)]*\\
& & \qquad
[(t(5,3)\;s(3,4)+t(5,6)\;s(6,4)]\}\\
M_{LL}^{c} & =& 2ig^2g_V^2 C_L^{Ve} \; C_L^{Ve}
\; D_f(p_2-p_4-p_3)
\; D_V(p_3+p_4) \; D_W(p_1-p_5) \\
& & \qquad t(2,3)\;s(1,6)*[ s(4,2)\;t(2,5) -s(4,3)\;t(3,5)] \\
M_{LR}^{c} & =& 2ig^2 g_V^2 C_L^{Ve} \; C_R^{Ve}
\; D_f(p_2-p_3-p_4)
\; D_V(p_3+p_4) \; D_W(p_1-p_5) \\
& & \qquad t(2,4)\;s(1,6)*[s(3,2)\;t(2,5)-s(3,4)\;t(4,5)] \\
M_{LL}^{c-2} & =& 2ig^2g_Z^2 C_L^{Z\nu} \; C_L^{Ze}
\; D_f(p_6+p_3+p_4)
\; D_{Z^0}(p_3+p_4) \; D_W(p_1-p_5) \\
& & \qquad t(2,5)\;s(4,6)*[s(1,6)\;t(6,3) +s(1,4)\;t(4,3)] \\
M_{LR}^{c-2} & =& 2ig^2 g_Z^2 C_L^{Z\nu} \; C_R^{Ze}
\; D_f(p_6+p_3+p_4)
\; D_{Z^0}(p_3+p_4) \; D_W(p_1-p_5) \\
& & \qquad t(2,5)\;s(3,6)*[s(1,6)\;t(6,4)+s(1,3)\;t(3,4)] \\
M_{LL}^{c-3} & =& -2ig^2 g_Z^2  C_L^{Z\nu} \; C_L^{Ze}
\; D_f(p_5+p_4+p_3)
\; D_{Z^0}(p_3+p_4) \; D_W(p_2-p_6) \\
& & \qquad t(5,3) \; s(6,1) * [ s(4,5) \; t(5,2) +s(4,3) \; t(3,2)] \\
M_{LR}^{c-3} & =& -2ig^2g_Z^2 C_L^{Z\nu} \; C_R^{Ze}
\; D_f(p_5+p_4+p_3)
\; D_{Z^0}(p_3+p_4) \; D_W(p_2-p_6) \\
& & \qquad t(5,4)\;s(6,1)*[s(3,5)t(5,2)+s(3,4)t(4,2)]\\
M_{LR}^{c-4} &=& -2ig^2g_V^2 C_L^{Ve} \; C_R^{Ve}
\; D_f(p_1 -p_3-p_4)
\; D_V(p_3+p_4)
\; D_W(p_2-p_6) \\
& & \qquad t(5,2)\;s(3,1)*[s(6,1)\;t(1,4)-s(6,3)\;t(3,4)] \\
M_{LL}^{c-4} &=& -2ig^2g_V^2 C_L^{Ve} \; C_L^{Ve}
\; D_f(p_1 -p_3-p_4)
\; D_V(p_3+p_4)
\; D_W(p_2-p_6) \\
& & \qquad t(5,2)\;s(4,1)*[s(6,1)\;t(1,3)-s(6,4)\;t(4,3)] \\
M_{LL}^{d} &=& {{-4ig^2g_V^2}\over{\cos^2\theta_w}}
C_L^{Z\nu}\;C_L^{Ze}\; C_L^{Ve} \; C_L^{Ve}
\; D_f(p_1 -p_3-p_4)
\; D_V(p_3+p_4)
\; D_Z(p_5+p_6) \\
& & \qquad t(2,5)\;s(4,1)*[s(6,1)\;t(1,3)-s(6,4)\;t(4,3)] \\
M_{LR}^{d} &=& {{-4ig^2g_V^2}\over{\cos^2\theta_w}}
C_L^{Z\nu}\; C_L^{Ze} \; C_L^{Ve} \; C_R^{Ve}
\; D_f(p_1 -p_3-p_4)
\; D_V(p_3+p_4)
\; D_Z(p_5+p_6) \\
& & \qquad t(2,5)\;s(3,1)*[s(6,1)\;t(1,4)-s(6,3)\;t(3,4)] \\
M_{RL}^{d} &=& {{-4ig^2g_V^2}\over{\cos^2\theta_w}}
C_L^{Z\nu}\; C_R^{Ze} \; C_R^{Ve} \; C_L^{Ve}
\; D_f(p_1 -p_3-p_4)
\; D_V(p_3+p_4)
\; D_Z(p_5+p_6) \\
& & \qquad s(2,6)\;t(3,1)*[t(5,1)\;s(1,4)-t(5,3)\;s(3,4)] \\
M_{RR}^{d} &=& {{-4ig^2g_V^2}\over{\cos^2\theta_w}}
C_L^{Z\nu}\; C_R^{Ze} \; C_R^{Ve} \; C_R^{Ve}
\; D_f(p_1 -p_3-p_4)
\; D_V(p_3+p_4)
\; D_Z(p_5+p_6) \\
& & \qquad s(2,6)\;t(4,1)*[t(5,1)\;s(1,3)-t(5,4)\;s(4,3)] \\
M_{LL}^{d-2} &=& {{-4ig^2g_V^2}\over{\cos^2\theta_w}}
C_L^{Z\nu}\;C_L^{Ze}\; C_L^{Ve} \; C_L^{Ve}
\; D_f(p_1 -p_5-p_6)
\; D_V(p_3+p_4)
\; D_Z(p_5+p_6) \\
& & \qquad s(6,1)\;t(2,3)*[t(1,5)\;s(4,1)-t(6,5)\;s(4,6)] \\
M_{LR}^{d-2} &=& {{-4ig^2g_V^2}\over{\cos^2\theta_w}}
C_L^{Z\nu}\;C_L^{Ze}\; C_L^{Ve} \; C_R^{Ve}
\; D_f(p_1 -p_5-p_6)
\; D_V(p_3+p_4)
\; D_Z(p_5+p_6) \\
& & \qquad s(6,1)\;t(2,4)*[t(1,5)\;s(3,1)-t(6,5)\;s(3,6)] \\
M_{RL}^{d-2} &=& {{-4ig^2g_V^2}\over{\cos^2\theta_w}}
C_L^{Z\nu}\;C_R^{Ze}\; C_R^{Ve} \; C_L^{Ve}
\; D_f(p_1 -p_5-p_6)
\; D_V(p_3+p_4)
\; D_Z(p_5+p_6) \\
& & \qquad t(5,1)\;s(2,4)*[s(1,6)*t(3,1)-s(5,6)\;t(3,5)] \\
M_{RR}^{d-2} &=& {{-4ig^2g_V^2}\over{\cos^2\theta_w}}
C_L^{Z\nu}\;C_R^{Ze}\; C_R^{Ve} \; C_R^{Ve}
\; D_f(p_1 -p_5-p_6)
\; D_V(p_3+p_4)
\; D_Z(p_5+p_6) \\
& & \qquad  t(5,1)\;s(2,3)*[s(1,6)\;t(4,1)-s(5,6)\;t(4,5)] \\
\end{eqnarray*}
\begin{eqnarray*}
M_{LL}^{e} &=& {{+4ig^2g_V^2}\over{\cos^2\theta_w}}
C_L^{Z\nu}\;C_L^{Ze}\; C_L^{Ve} \; C_L^{Ve}
\; D_f(p_4+p_5+p_6)
\; D_V(p_1+p_2)
\; D_Z(p_5+p_6) \\
& & \qquad t(3,2)\;s(6,4)*[s(1,4)\;t(4,5)+s(1,6)\;t(6,5)] \\
M_{RL}^{e} &=& {{+4ig^2g_V^2}\over{\cos^2\theta_w}}
C_L^{Z\nu}\;C_L^{Ze}\; C_R^{Ve} \; C_L^{Ve}
\; D_f(p_4+p_5+p_6)
\; D_V(p_1+p_2)
\; D_Z(p_5+p_6) \\
& & \qquad t(3,1)\;s(6,4)*[s(2,4)\;t(4,5)+s(2,6)\;t(6,5)] \\
M_{LR}^{e} &=& {{+4ig^2g_V^2}\over{\cos^2\theta_w}}
C_L^{Z\nu}\;C_R^{Ze}\; C_R^{Ve} \; C_L^{Ve}
\; D_f(p_4+p_5+p_6)
\; D_V(p_1+p_2)
\; D_Z(p_5+p_6) \\
& & \qquad s(3,1)\;t(5,4)*[t(2,4)\;s(4,6)+t(2,5)\;s(5,6)] \\
M_{RR}^{e} &=& {{+4ig^2g_V^2}\over{\cos^2\theta_w}}
C_L^{Z\nu}\;C_R^{Ze}\; C_R^{Ve} \; C_R^{Ve}
\; D_f(p_4+p_5+p_6)
\; D_V(p_1+p_2)
\; D_Z(p_5+p_6) \\
& & \qquad s(3,2)\;t(5,4)*[t(1,4)\;s(4,6)+t(1,5)\;s(5,6)] \\
M_{LL}^{e-2} &=& {{-4ig^2g_V^2}\over{\cos^2\theta_w}}
C_L^{Z\nu}\;C_L^{Ze}\; C_L^{Ve} \; C_L^{Ve}
\; D_f(p_3+p_5+p_6)
\; D_V(p_1+p_2)
\; D_Z(p_5+p_6) \\
& & \qquad t(3,5)\;s(1,4)*[s(6,3)\;t(3,2)+s(6,5)\;t(5,2)] \\
M_{RL}^{e-2} &=& {{-4ig^2g_V^2}\over{\cos^2\theta_w}}
C_L^{Z\nu}\;C_L^{Ze}\; C_R^{Ve} \; C_L^{Ve}
\; D_f(p_3+p_5+p_6)
\; D_V(p_1+p_2)
\; D_Z(p_5+p_6) \\
& & \qquad t(3,5)\;s(2,4)*[s(6,3)\;t(3,1)+s(6,5)\;t(5,1)] \\
M_{LR}^{e-2} &=& {{-4ig^2g_V^2}\over{\cos^2\theta_w}}
C_L^{Z\nu}\;C_R^{Ze}\; C_R^{Ve} \; C_L^{Ve}
\; D_f(p_3+p_5+p_6)
\; D_V(p_1+p_2)
\; D_Z(p_5+p_6) \\
& & \qquad s(3,6)\;t(2,4)*[t(5,3)\;s(3,1)+t(5,6)\;s(6,1)] \\
M_{RR}^{e-2} &=& {{-4ig^2g_V^2}\over{\cos^2\theta_w}}
C_L^{Z\nu}\;C_R^{Ze}\; C_R^{Ve} \; C_R^{Ve}
\; D_f(p_3+p_5+p_6)
\; D_V(p_1+p_2)
\; D_Z(p_5+p_6) \\
& & \qquad s(3,6)\;t(1,4)*[t(5,3)\;s(3,2)+t(5,6)\;s(6,2)] \\
M_{LL}^{f} &=& {+4ig_Z^4}
{C_L^{Z\nu}}^2\; {C_L^{Ze}}^2
\; D_f(p_3+p_4+p_6)
\; D_Z(p_1+p_2)
\; D_Z(p_3+p_4) \\
& & \qquad t(5,2)\;s(4,6)*[s(1,6)\;t(6,3)+s(1,4)\;t(4,3)]\\
M_{LR}^{f} &=& {+4ig_Z^4}
{C_L^{Z\nu}}^2\; {C_L^{Ze}}\; C_R^{Ze}
\; D_f(p_3+p_4+p_6)
\; D_Z(p_1+p_2)
\; D_Z(p_3+p_4) \\
& & \qquad t(5,2)\;s(3,6)*[s(1,6)\;t(6,4)+s(1,3)\;t(3,4)]\\
M_{RL}^{f} &=& {+4ig_Z^4}
{C_L^{Z\nu}}^2\; {C_L^{Ze}}\; C_R^{Ze}
\; D_f(p_3+p_4+p_6)
\; D_Z(p_1+p_2)
\; D_Z(p_3+p_4) \\
& & \qquad t(5,1)\;s(4,6)*[s(2,6)\;t(6,3)+s(2,4)\;t(4,3)]\\
M_{RR}^{f} &=& {+4ig_Z^4}
{C_L^{Z\nu}}^2\; {C_L^{Ze}}^2
\; D_f(p_3+p_4+p_6)
\; D_Z(p_1+p_2)
\; D_Z(p_3+p_4) \\
& & \qquad t(5,1)\;s(3,6)*[s(2,6)\;t(6,4)+s(2,3)\;t(3,4)]\\
M_{LL}^{f-2} &=& {-4ig_Z^4}
{C_L^{Z\nu}}^2\;{C_L^{Ze}}^2
\; D_f(p_3+p_4+p_5)
\; D_Z(p_1+p_2)
\; D_Z(p_3+p_4) \\
& & \qquad t(5,3)\;s(1,6)*[s(4,5)\;t(5,2)+s(4,3)\;t(3,2)] \\
M_{LR}^{f-2} &=& {-4ig_Z^4}
{C_L^{Z\nu}}^2\; C_L^{Ze} \; C_R^{Ze}
\; D_f(p_3+p_4+p_5)
\; D_Z(p_1+p_2)
\; D_Z(p_3+p_4) \\
& & \qquad t(5,4)\;s(1,6)*[s(3,5)\;t(5,2)+s(3,4)\;t(4,2)] \\
M_{RL}^{f-2} &=& {-4ig_Z^4}
{C_L^{Z\nu}}^2\; C_L^{Ze} \; C_R^{Ze}
\; D_f(p_3+p_4+p_5)
\; D_Z(p_1+p_2)
\; D_Z(p_3+p_4) \\
& & \qquad t(5,3)\;s(2,6)*[s(4,5)\;t(5,1)+s(4,3)\;t(3,1)] \\
M_{RR}^{f-2} &=& {-4ig_Z^4}
{C_L^{Z\nu}}^2\; {C_R^{Ze}}^2
\; D_f(p_3+p_4+p_5)
\; D_Z(p_1+p_2)
\; D_Z(p_3+p_4) \\
& & \qquad t(5,4)\;s(2,6)*[s(3,5)\;t(5,1)+s(3,4)\;t(4,1)] \\
M_{LL}^{g} &=& {-ig^4}
\; D_f(p_1-p_3-p_6)
\; D_W(p_3+p_6)
\; D_W(p_4+p_5) \\
& & \qquad t(2,5)\;s(6,1)*[s(4,1)\;t(1,3)-s(4,6)\;t(6,3)]\\
M_{LL}^{h} &=& {-ig^4}
\; D_f(p_3-p_1+p_5)
\; D_W(p_1-p_5)
\; D_W(p_2-p_6) \\
& & \qquad t(3,5)\;s(6,4)*[s(1,3)\;t(3,2)+s(1,5)\;t(5,2)] \\
M_{LL}^{i} &=& -2ig^2 g_V^2
\; C_L^{Ze} \; C_L^{Ze}
\; D_f(p_3+p_5+p_6)
\; D_V(p_1+p_2)
\; D_W(p_3+p_6) \\
& & \qquad t(5,3)\;s(1,4)*[s(6,5)\;t(5,2)+s(6,3)\;t(3,2)] \\
M_{RL}^{i} &=& -2ig^2 g_V^2
\; C_R^{Ze} \; C_L^{Ze}
\; D_f(p_3+p_5+p_6)
\; D_V(p_1+p_2)
\; D_W(p_3+p_6) \\
& & \qquad t(5,3)\;s(2,4)*[s(6,5)\;t(5,1)+s(6,3)\;t(3,1)] \\
M_{LL}^{i-2} &=& +2ig^2 g_V^2
\; C_L^{Ze} \; C_L^{Ze}
\; D_f(p_4+p_5+p_6)
\; D_V(p_1+p_2)
\; D_W(p_4+p_5) \\
& & \qquad t(3,2)\;s(4,6)*[s(1,4)\;t(4,5)+s(1,6)\;t(6,5)] \\
M_{RL}^{i-2} &=& +2ig^2 g_V^2
\; C_R^{Ze} \; C_L^{Ze}
\; D_f(p_4+p_5+p_6)
\; D_V(p_1+p_2)
\; D_W(p_4+p_5) \\
& & \qquad t(3,1)\;s(4,6)*[s(2,4)\;t(4,5)+s(2,6)\;t(6,5)] \\
M_{LL}^{j} &=& +2ig^2 g_Z^2
\; C_L^{Z\nu} \; C_L^{Ze}
\; D_f(p_3+p_4+p_6)
\; D_Z(p_1+p_2)
\; D_W(p_3+p_6) \\
& & \qquad t(5,2)\;s(6,4)*[s(1,6)\;t(6,3)+s(1,4)\;t(4,3)] \\
M_{RL}^{j} &=& +2ig^2 g_Z^2
\; C_L^{Z\nu} \; C_R^{Ze}
\; D_f(p_3+p_4+p_6)
\; D_Z(p_1+p_2)
\; D_W(p_3+p_6) \\
& & \qquad t(5,1)\;s(6,4)*[s(2,6)\;t(6,3)+s(2,4)\;t(4,3)] \\
M_{LL}^{j-2} &=& -2ig^2 g_Z^2
\; C_L^{Z\nu} \; C_L^{Ze}
\; D_f(p_3+p_4+p_5)
\; D_Z(p_1+p_2)
\; D_W(p_4+p_5) \\
& & \qquad t(3,5)\;s(1,6)*[s(4,5)\;t(5,2)+s(4,3)\;t(3,2)] \\
M_{RL}^{j-2} &=& -2ig^2 g_Z^2
\; C_L^{Z\nu} \; C_R^{Ze}
\; D_f(p_3+p_4+p_5)
\; D_Z(p_1+p_2)
\; D_W(p_4+p_5) \\
& & \qquad t(3,5)\;s(2,6)*[s(4,5)\;t(5,1)+s(4,3)\;t(3,1)] \\
\end{eqnarray*}

where $c-2$ is obtained from $c$ with the $Z$ coming from the $\bar{\nu_e}$
leg; $c-3$ is obtained from $c$ with the $Z$ coming from the  $\nu_e$ leg;
$c-4$ is obtained from $c$ with the $\gamma, Z$ coming from the $e^-$ leg;
$d-2$ is obtained from $d$ with the $Z$ coming from the $e^-$ leg; $e-2$ is
obtained from $e$ with the $Z$ coming from the $\mu^-$ leg; $f-2$ is obtained
from $f$ with the $Z$ coming from the $\nu$ leg; $i-2$ is obtained from $i$
with the $W$ coming from the $\mu^+$ leg; $j-2$ is obtained from $j$ with the
$W$ coming from the $\nu$ leg.

The propagator denominators are defined as
\begin{eqnarray*}
D_f(p_i) & = & (p_i^2)^{-1} \\
D_W(p_i) & = & ( p_i^2  - M_W^2 +i \Gamma_W M_W )^{-1} \\
D_\gamma(p_i) & = & (p_i^2)^{-1} \\
D_{Z^0}(p_i) & = & ( p_i^2 - M_Z^2 +i \Gamma_Z M_Z )^{-1}\\
\end{eqnarray*}

and $C_R^e=\sin^2\theta_W$ and $C_L^e = -{1\over 2} +\sin^2 \theta_W$.
In diagrams (a) and (b) $p_Z$ and $p_{W^\pm}$ represent the
gauge boson momentum flowing {\it into} the vertex.
The first subscript of the amplitudes refers to the helicity of the
electron and the second subscript to the helicity of the muon.
To obtain the cross section the amplitudes for given electron
and photon helicities are summed over and squared.  These are then
averaged to obtain the spin averaged matrix element squared and
finally integrated over the final state phase space to yield the cross section.

\figure{The trilinear Gauge Boson Vertex
\label{vertex}}

\figure{The Feynman diagrams contributing to the process
$e^+e^- \to \mu^+ \mu^- \nu \bar{\nu}$
\label{diagrams}}

\figure{The cross section $\sigma(e^+e^- \to \mu^+ \mu^- \nu
\bar{\nu})$
as a function of $\sqrt{s}$.  The solid line is for the cuts on $E_{\mu^\pm}$,
$\not{p}_T$, and $\cos\theta_{\mu^\pm}$ given in the text.  The dashed line
adds the cut $M_{\mu^+\mu^-}> 25$~GeV and the dotted line has
$|M_{\mu^+\mu^-}-M_Z| < 5$~GeV. }

\figure{(a) $d\sigma/dM_{\mu^+\mu-}$ and (b)
$d\sigma/d\cos\theta_{\mu^+\mu^-}$ at $\sqrt{s}=200$~GeV.  In both cases the
solid line is for standard model values of $\kappa_\gamma$, $\lambda_\gamma$,
$\kappa_Z$, and $\lambda_Z$,
the long dashed line is for $\delta\kappa_\gamma=\lambda_\gamma=\lambda_Z=0$
and $\delta\kappa_Z=2$;
the dotted line is for $\delta\kappa_\gamma=\lambda_\gamma=\delta\kappa_Z=0$;
and $\lambda_Z=2$
the dot-dashed line is for $\delta\kappa_Z=\lambda_\gamma=\lambda_Z=0$
and $\delta\kappa_\gamma=2$
and dot-dot-dashed line is for $\delta\kappa_\gamma=
\lambda_Z=\delta\kappa_Z=0$ and $\lambda_\gamma=2$ where
$\delta\kappa_V=\kappa_V-1$.  In this figure and all subsequent ones, the
small bumps are to due statistical fluctuations arising from the Monte-Carlo
phase space integration.}

\figure{Sensitivities of the TGV's to anomalous couplings at 95\% C.L.
based on the kinematic cuts given in the text.
The solid lines are based on the statistics assuming an integrated luminosity
of 250~pb$^{-1}$ and the dashed lines are based on integrated luminosities of
500~pb$^{-1}$.}

\figure{$d\sigma/dM_{\mu^+\mu-}$ at $\sqrt{s}=500$~GeV.  The
solid line is for standard model values of $\kappa_\gamma$, $\lambda_\gamma$,
$\kappa_Z$, and $\lambda_Z$,
the long dashed line is for $\delta\kappa_\gamma=\lambda_\gamma=\lambda_Z=0$
and $\delta\kappa_Z=-0.5$;
the dotted line is for $\delta\kappa_\gamma=\lambda_\gamma=\delta\kappa_Z=0$
and $\lambda_Z=1$
the dot-dashed line is for $\delta\kappa_Z=\lambda_\gamma=\lambda_Z=0$
and $\delta\kappa_\gamma=0.5$
and dot-dot-dashed line is for $\delta\kappa_\gamma=
\lambda_Z=\delta\kappa_Z=0$ and $\lambda_\gamma=0.5$. }

\figure{(a) $d\sigma/d\cos\theta_{\mu^+\mu^-}$
and (b) $d\sigma/d{p_T}_Z$ at $\sqrt{s}=500$~GeV
with $|M_{\mu^+\mu^-}-M_Z| < 5$~GeV.
In both cases the
solid line is for standard model values of $\kappa_\gamma$, $\lambda_\gamma$,
$\kappa_Z$, and $\lambda_Z$,
the long dashed line is for $\delta\kappa_\gamma=\lambda_\gamma=\lambda_Z=0$
and $\delta\kappa_Z=-0.5$;
the dotted line is for $\delta\kappa_\gamma=\lambda_\gamma=\delta\kappa_Z=0$
and $\lambda_Z=0.5$
and dot-dashed line is for $\delta\kappa_\gamma=
\lambda_Z=\delta\kappa_Z=0$ and $\lambda_\gamma=-0.5$.}

\figure{Sensitivities of the TGV's to anomalous couplings
for $\sqrt{s}=500$~GeV and L=10~fb$^{-1}$
based on (a) $d\sigma/d\cos\theta_{\mu^+\mu^-}$
(b) $d\sigma/d{p_T}_Z$
with $|M_{\mu^+\mu^-}-M_Z| < 5$~GeV using the binning given in the text.
In both cases the solid lines are 68\% C.L., the dashed lines are 90\% C.L.,
and the dot-dashed curves are 95\% C.L..}

\figure{95 \% C.L. bounds of the TGV's
based on (a) $d\sigma/d\cos\theta_{\mu^+\mu^-}$
(b) $d\sigma/d{p_T}_Z$ at $\sqrt{s}=500$~GeV
with $|M_{\mu^+\mu^-}-M_Z| < 5$~GeV using the binning given in the text.
In both cases the solid curves are based on 10~fb$^{-1}$,
the dashed curves on 50~fb$^{-1}$, the dot-dashed curves on
10~fb$^{-1}$ + $\delta^{sys}$, and
the dotted curves on 50~fb$^{-1}$ + $\delta^{sys}$ where
$\delta^{sys}=$5\%.}

\figure{Sensitivities of the TGV's to anomalous couplings
for $\sqrt{s}=500$~GeV and L=10~fb$^{-1}$
based on the total cross section integrated over the kinematic region
$110\leq M_{\mu^+\mu^-}\leq 400$ GeV.
The solid lines are 68\% C.L., the dashed lines are 90\% C.L.,
and the dot-dashed curves are 95\% C.L..}

\figure{Sensitivities of the TGV's to anomalous couplings
for $\sqrt{s}=1$~TeV and L=50~fb$^{-1}$
based on (a) $d\sigma/d\cos\theta_{\mu^+\mu^-}$
(b) $d\sigma/d{p_T}_Z$
with $|M_{\mu^+\mu^-}-M_Z| < 5$~GeV using the binning given in the text.
In both cases the solid lines are 68\% C.L., the dashed lines are 90\% C.L.,
and the dot-dashed curves are 95\% C.L..}

\figure{$d\sigma/dM_{\mu^+\mu-}$ at $\sqrt{s}=1$~TeV.  The
solid line is for standard model values of $\kappa_\gamma$, $\lambda_\gamma$,
$\kappa_Z$, and $\lambda_Z$,
the long dashed line is for $\delta\kappa_\gamma=\lambda_\gamma=\lambda_Z=0$
and $\delta\kappa_Z=-0.2$;
the dotted line is for $\delta\kappa_Z=\lambda_\gamma=\lambda_Z=0$
and $\delta\kappa_\gamma=0.2$
and dot-dashed line is for $\delta\kappa_\gamma=
\lambda_Z=\delta\kappa_Z=0$ and $\lambda_\gamma=-0.2$. }

\figure{Sensitivities of the TGV's to anomalous couplings
for $\sqrt{s}=1$~TeV and L=50~fb$^{-1}$
based on the total cross section integrated over the kinematic region
$200\leq M_{\mu^+\mu^-}\leq 900$ GeV.
The solid lines are 68\% C.L., the dashed lines are 90\% C.L.,
and the dot-dashed curves are 95\% C.L..}

\newpage
\begin{table}
\caption{Sensitivities to $\kappa_\gamma$, $\lambda_\gamma$
$\kappa_Z$, and $\lambda_Z$ at 95\% C.L.
from the process $e^+e^- \to \mu^+ \mu^- \nu
\bar{\nu}$ at a 500~GeV $e^+e^-$ collider.
The statistical error is based on the specified integrated luminosity
and $\delta^{sys}$ refers to the systematic error which we take as 5\%.}
\begin{tabular}{lllll}
\multicolumn{5}{c}{Based on $\sigma(e^+e^- \to \mu^+ \mu^- \nu \bar{\nu})$
with $|M_{\mu^+\mu^-}-M_Z| < 5$~GeV} \\
\tableline
	& L=20~fb$^{-1}$ & L=50~fb$^{-1}$
	& L=20~fb$^{-1}$ $+\delta^{sys}$
	& L=50~fb$^{-1}$ $+\delta^{sys}$ \\
\tableline
$\delta\kappa_Z$ & $\pm 0.1$ & $\pm 0.06$ & ${+0.25\atop -0.30}$
	& ${+0.24\atop -0.28}$ \\
$\delta\lambda_Z$ & ${+0.18\atop -0.53}$ & ${+0.12\atop -0.48}$
	& ${+0.34\atop -0.7}$ & ${+0.33\atop -0.7}$ \\
\tableline
\multicolumn{5}{c}{Based on $d\sigma/d{p_T}Z$
	with $|M_{\mu^+\mu^-}-M_Z| < 5$~GeV} \\
\tableline
$\delta\kappa_Z$ & ${+0.19\atop -0.23}$ & $\pm 0.09$
	& ${+0.26\atop -0.32}$ 	& ${+0.19\atop -0.23}$ \\
$\delta\lambda_Z$ & ${+0.27\atop -0.36}$ & ${+0.15\atop -0.20}$
	& ${+0.33\atop -0.43}$ & ${+0.27\atop -0.38}$ \\
\tableline
\multicolumn{5}{c}{Based on binning  $M_{\mu^+\mu^-}$} \\
\tableline
	& L=10~fb$^{-1}$ & L=50~fb$^{-1}$
	& L=10~fb$^{-1}$ $+\delta^{sys}$
	& L=50~fb$^{-1}$ $+\delta^{sys}$ \\
\tableline
$\delta\kappa_Z$ & ${+0.13\atop -0.09}$ & ${+0.08\atop -0.05}$
	& ${+0.15\atop -0.12}$ 	& ${+0.13\atop -0.10}$ \\
$\delta\lambda_Z$ & ${+0.082\atop -0.090}$ & ${+0.055\atop -0.060}$
	& ${+0.096\atop -0.107}$ & ${+0.082\atop -0.090}$ \\
$\delta\kappa_\gamma$ & ${+0.21\atop -0.09}$ & ${+0.17\atop -0.05}$
	& ${+0.27\atop -0.11}$ 	& ${+0.21\atop -0.08}$ \\
$\delta\lambda_\gamma$ & ${+0.09\atop -0.12}$ & ${+0.06\atop -0.07}$
	& ${+0.12\atop -0.14}$ & ${+0.09\atop -0.12}$ \\
\end{tabular}
\end{table}

\newpage
\begin{table}
\caption{Sensitivities to $\kappa_\gamma$, $\lambda_\gamma$
$\kappa_Z$, and $\lambda_Z$ at 95\% C.L.
from the process $e^+e^- \to \mu^+ \mu^- \nu
\bar{\nu}$ at a 1~TeV $e^+e^-$ collider.
The statistical error is based on the specified integrated luminosity
and $\delta^{sys}$ refers to the systematic error which we take as 5\%.}
\begin{tabular}{lllll}
	& L=50~fb$^{-1}$ & L=200~fb$^{-1}$
	& L=50~fb$^{-1}$ $+\delta^{sys}$
	& L=200~fb$^{-1}$ $+\delta^{sys}$ \\
\tableline
\multicolumn{5}{c}{Based on $d\sigma/d\cos\theta_{\mu^+\mu^-}$
	with $|M_{\mu^+\mu^-}-M_Z| < 5$~GeV} \\
\tableline
$\delta\kappa_Z$ & $\pm 0.07$ & $\pm 0.03$
	& ${+0.13\atop -0.16}$ 	& ${+0.11\atop -0.13}$ \\
$\delta\lambda_Z$ & ${+0.07\atop -0.14}$ & ${+0.04\atop -0.06}$
	& ${+0.12\atop -0.25}$ & ${+0.10\atop -0.16}$ \\
\tableline
\multicolumn{5}{c}{Based on $d\sigma/d{p_T}Z$
	with $|M_{\mu^+\mu^-}-M_Z| < 5$~GeV} \\
\tableline
$\delta\kappa_Z$ & $\pm 0.065$ & $\pm 0.03$
	& ${+0.17\atop -0.19}$ 	& ${+0.16\atop -0.18}$ \\
$\delta\lambda_Z$ & ${+0.08\atop -0.10}$ & ${+0.040\atop -0.055}$
	& $\pm 0.18$ & $ \pm 0.17$ \\
\tableline
\multicolumn{5}{c}{Based on  $200<M_{\mu^+\mu^-}<900$~GeV} \\
\tableline
$\delta\kappa_Z$ & ${+0.030\atop -0.012}$ & ${+0.025\atop -0.007}$
	& ${+0.038\atop -0.018}$ 	& ${+0.034\atop -0.016}$ \\
$\delta\lambda_Z$ & ${+0.021\atop -0.011}$ & ${+0.017\atop -0.007}$
	& ${+0.026\atop -0.025}$ & $\pm 0.024$ \\
$\delta\kappa_\gamma$ & ${+0.044\atop -0.011}$ & ${+0.008\atop -0.007}$
	& ${+0.048\atop -0.016}$ 	& ${+0.046\atop -0.014}$ \\
$\delta\lambda_\gamma$ & ${+0.029\atop -0.009}$ & ${+0.025\atop -0.006}$
	& ${+0.033\atop -0.014}$ & ${+0.032\atop -0.012}$ \\
$\delta\kappa_Z$ & ${+0.030\atop -0.012}$ & ${+0.025\atop -0.007}$
	& ${+0.038\atop -0.018}$ 	& ${+0.034\atop -0.016}$ \\
$\delta\lambda_Z$ & ${+0.021\atop -0.011}$ & ${+0.017\atop -0.007}$
	& ${+0.026\atop -0.025}$ & $\pm 0.024$ \\
$\delta\kappa_\gamma=\delta\kappa_Z
	$ & ${+0.027\atop -0.006}$ & ${+0.006\atop -0.003}$
	& ${+0.033\atop -0.009}$ 	& ${+0.032\atop -0.008}$ \\
$\delta\lambda_\gamma=\delta\lambda_Z$
	& ${+0.008\atop -0.006}$ & $\pm 0.004$
	& ${+0.022\atop -0.009}$ & ${+0.021\atop -0.008}$ \\
\end{tabular}
\end{table}

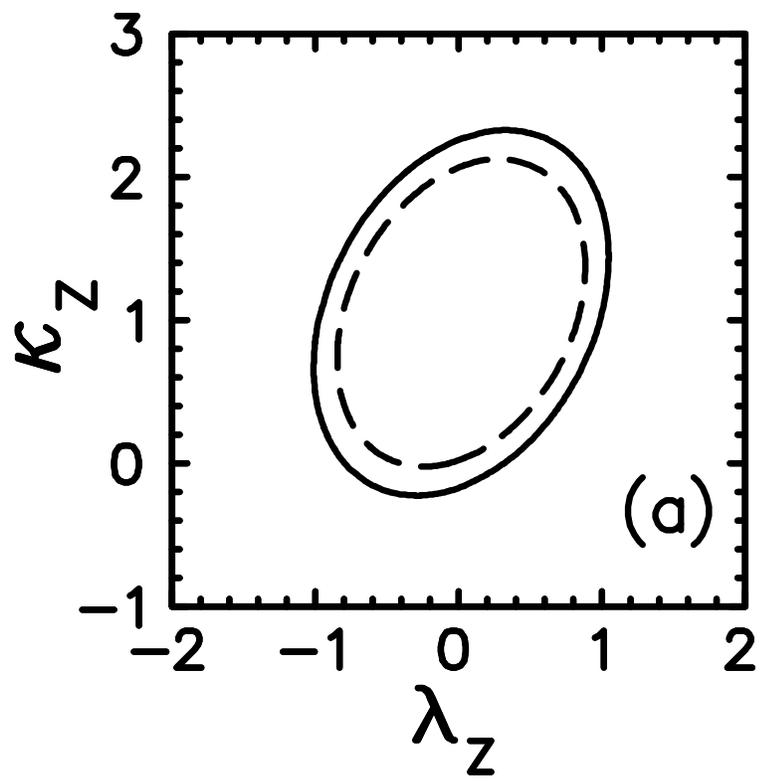
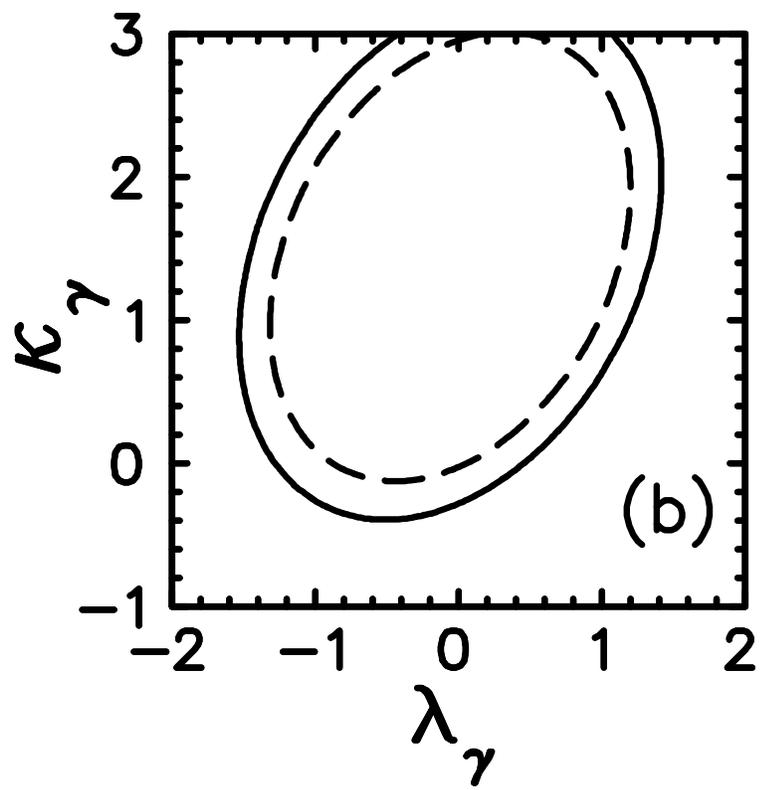
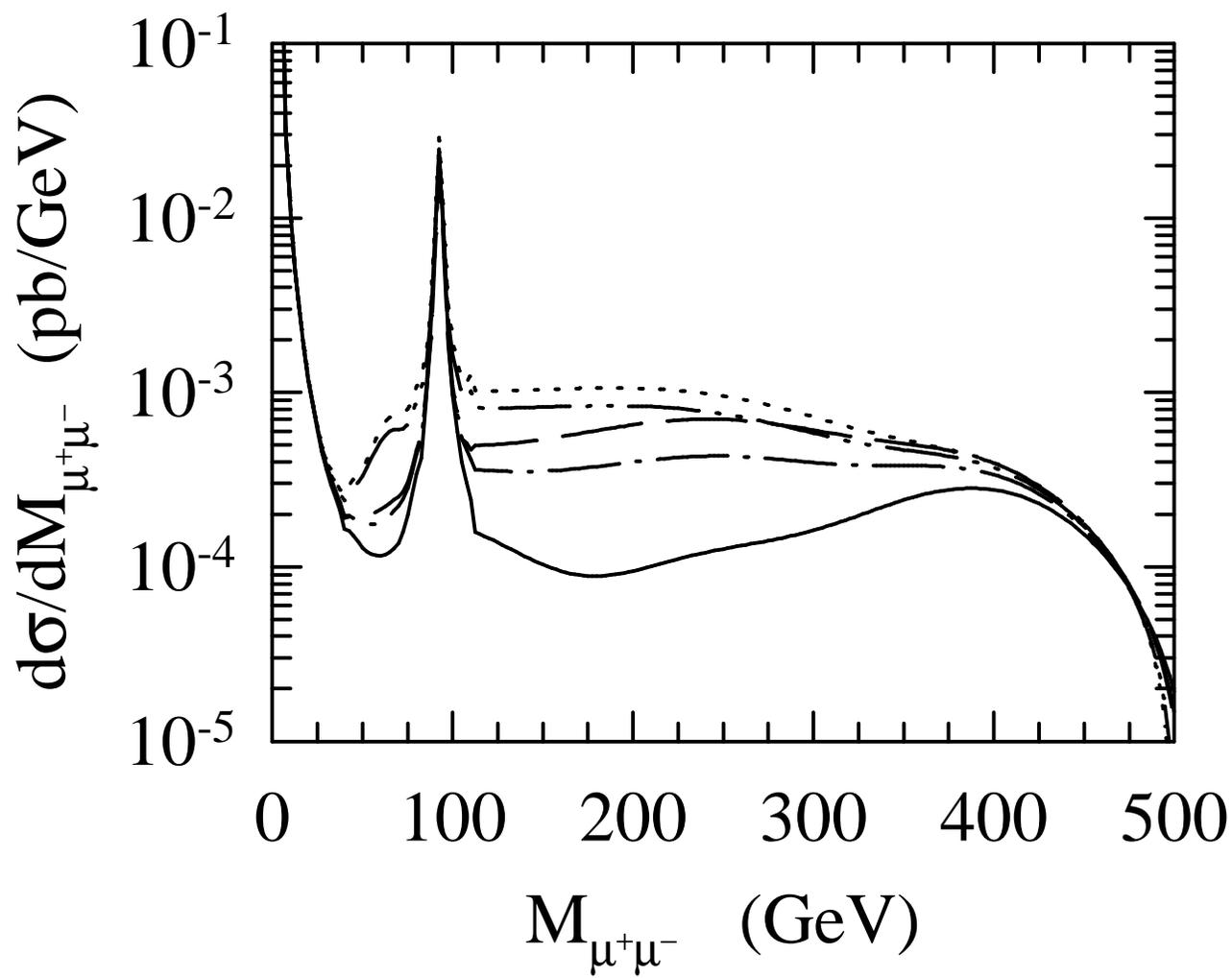
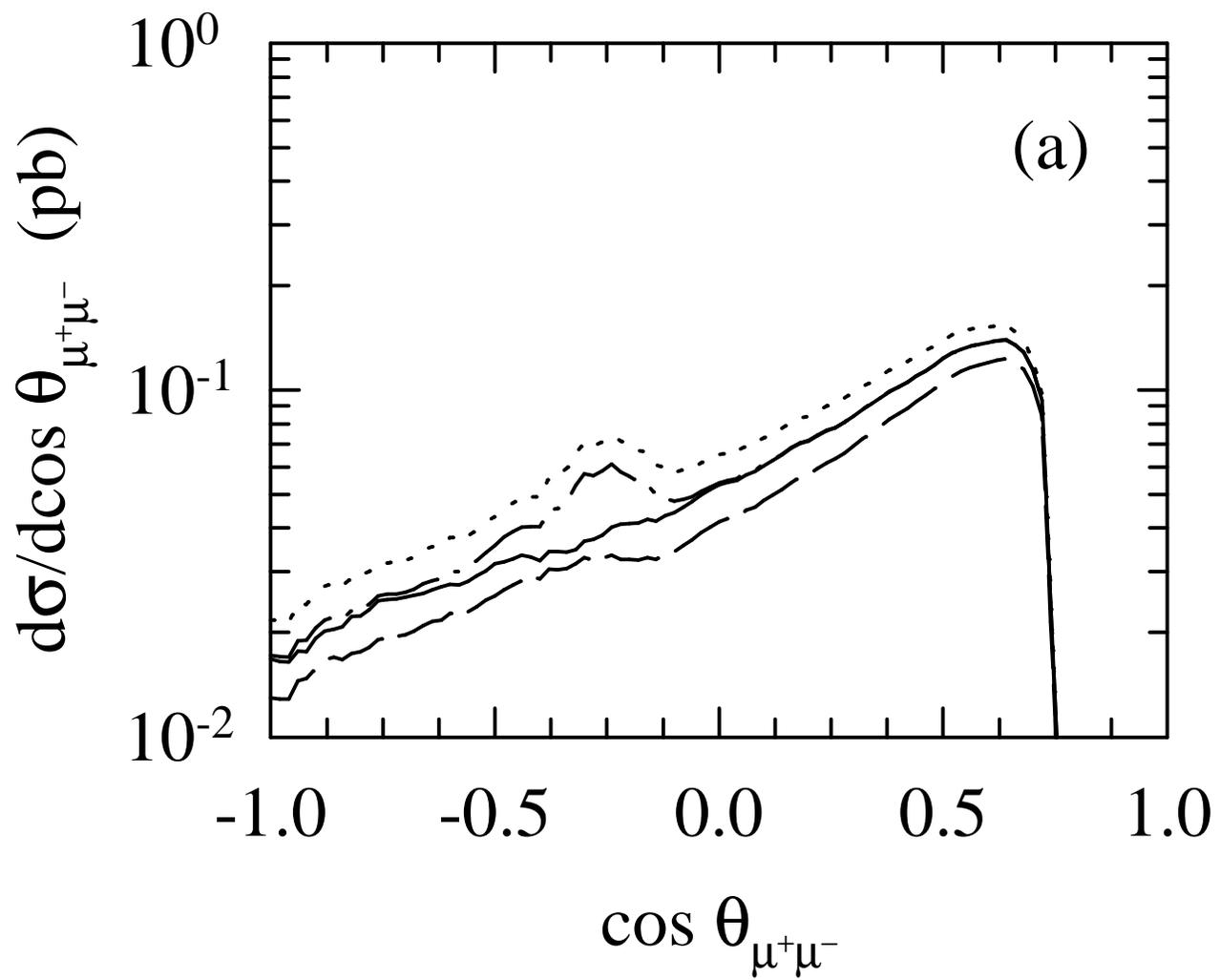
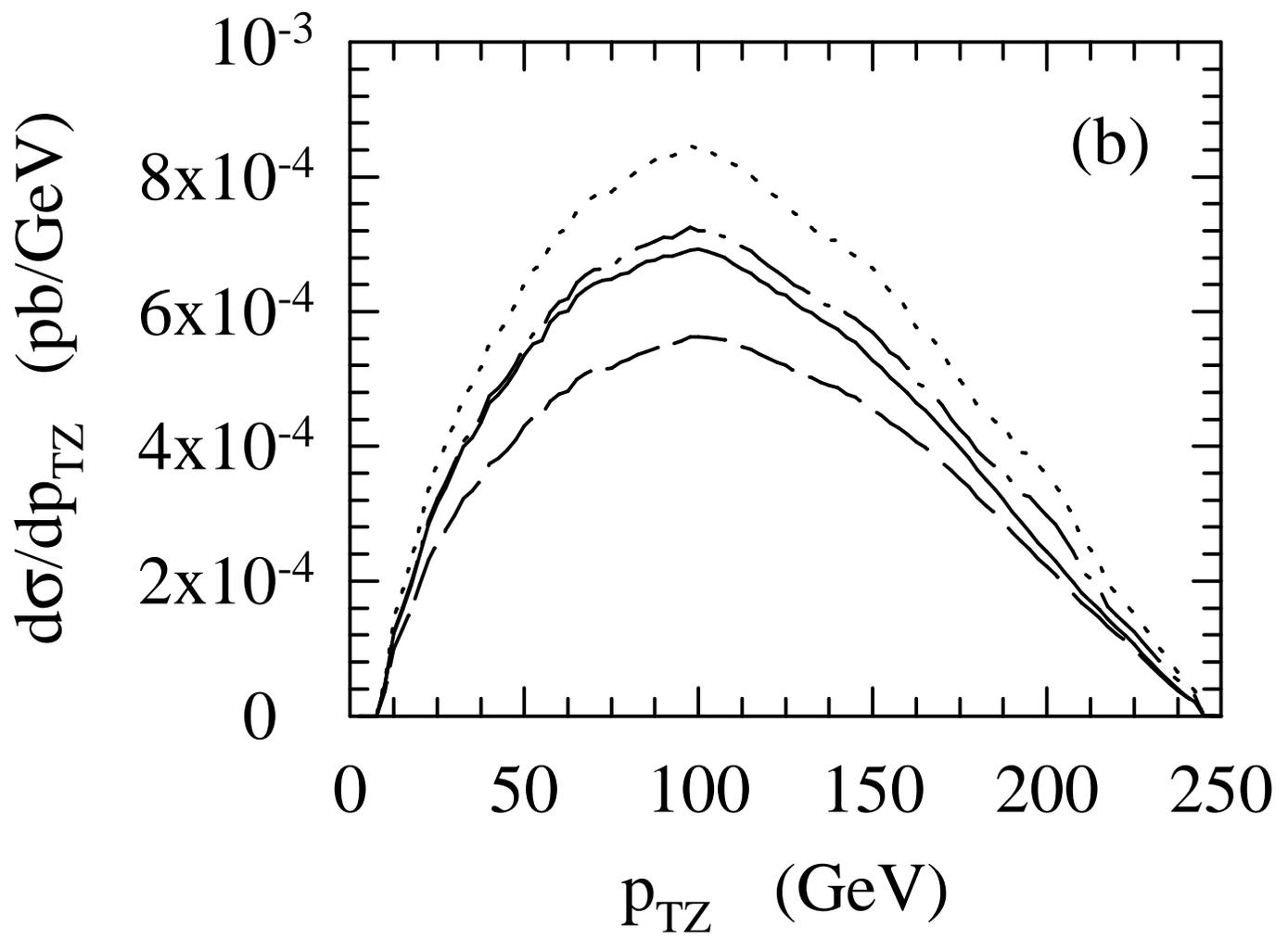
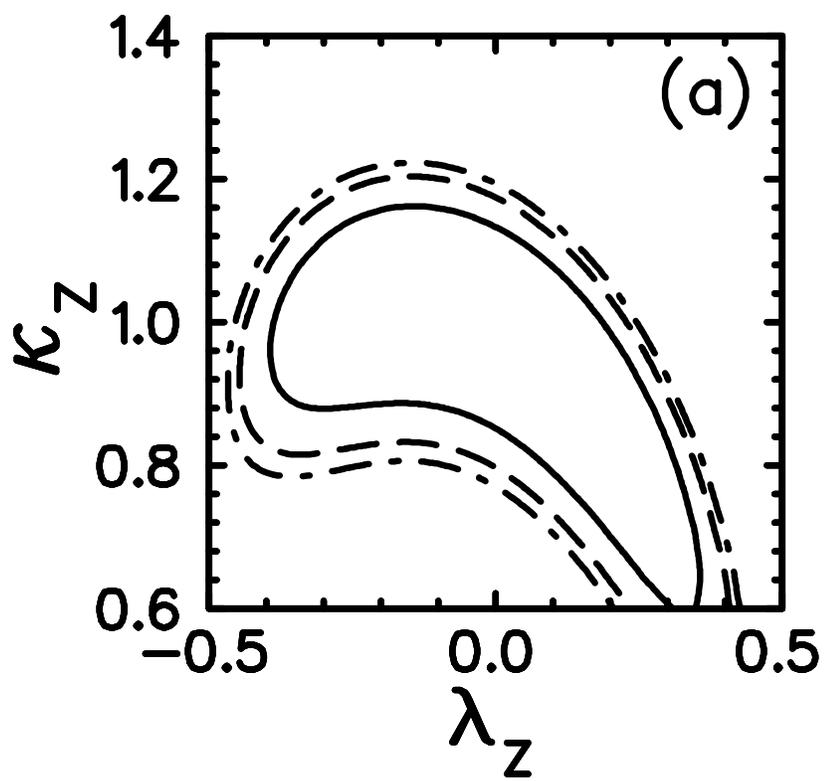
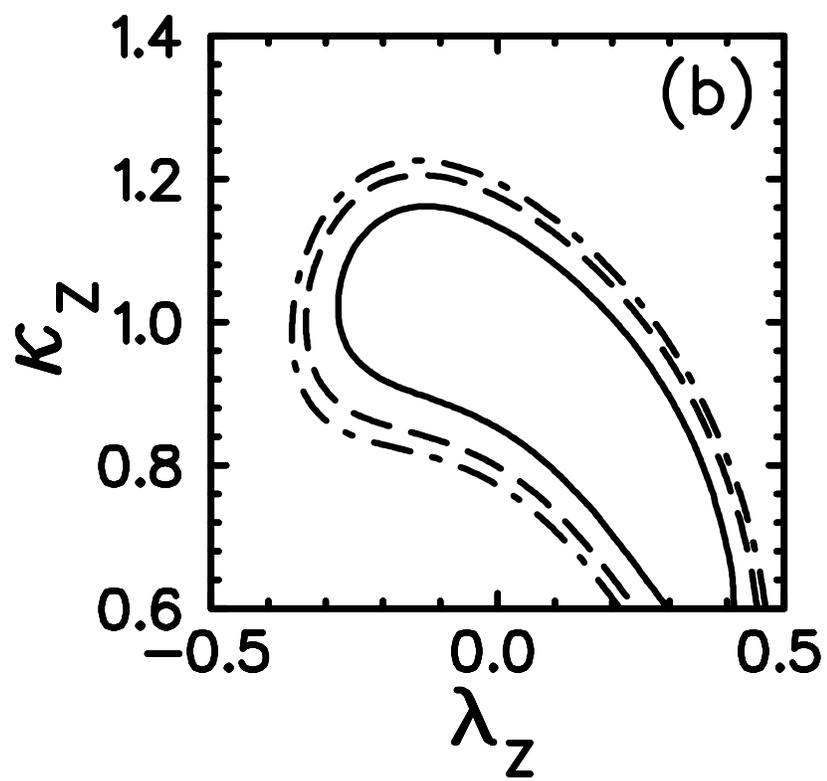
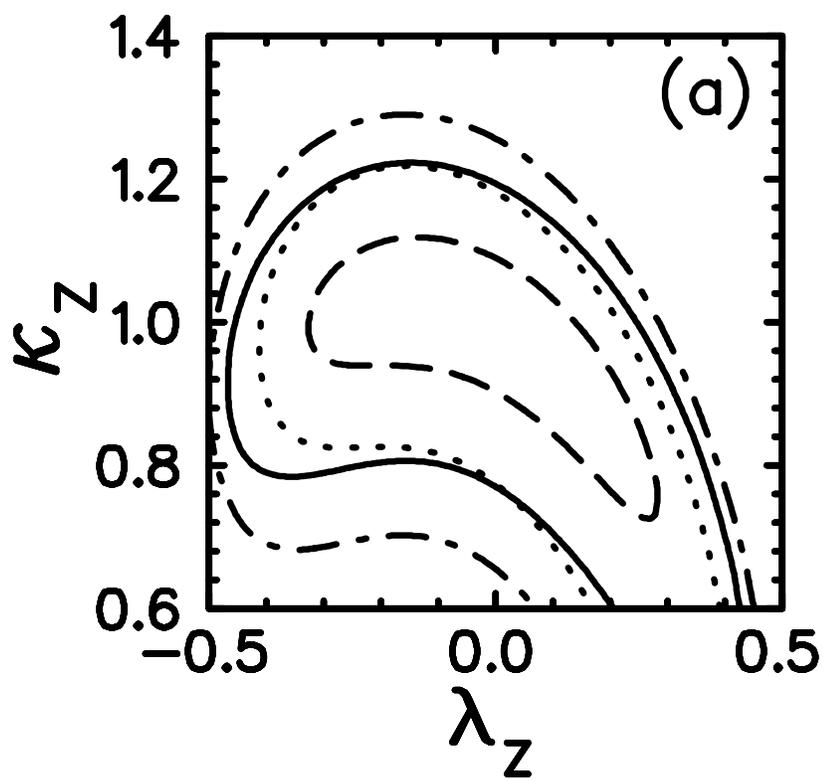
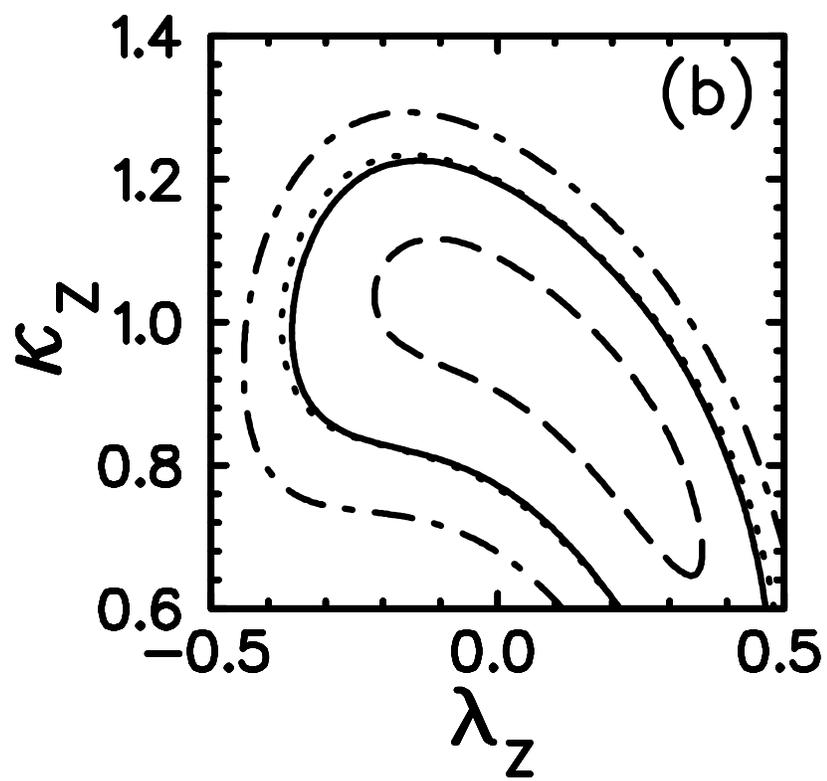
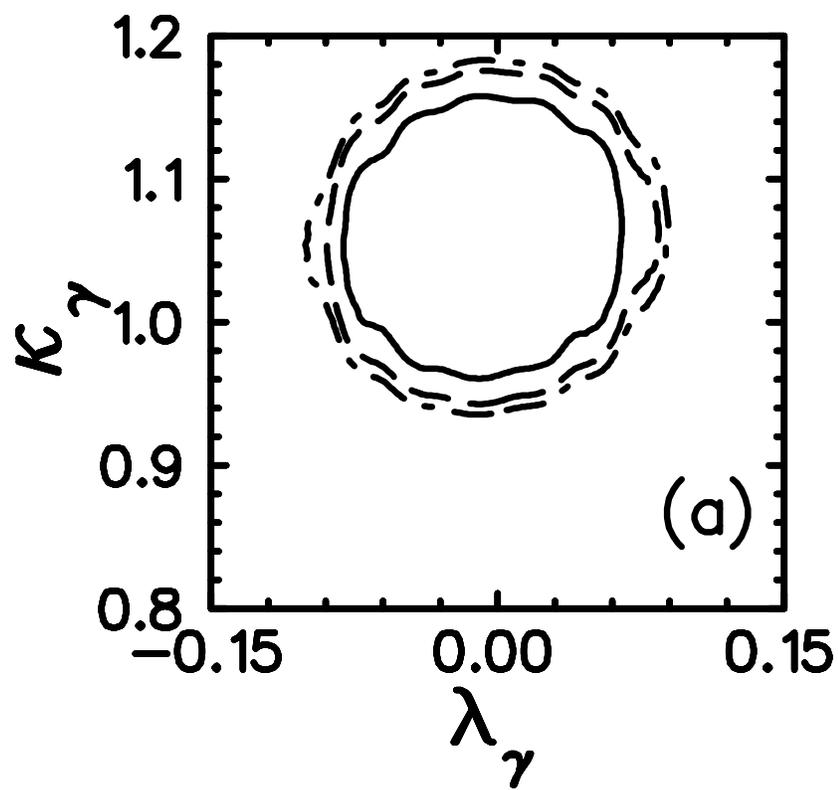
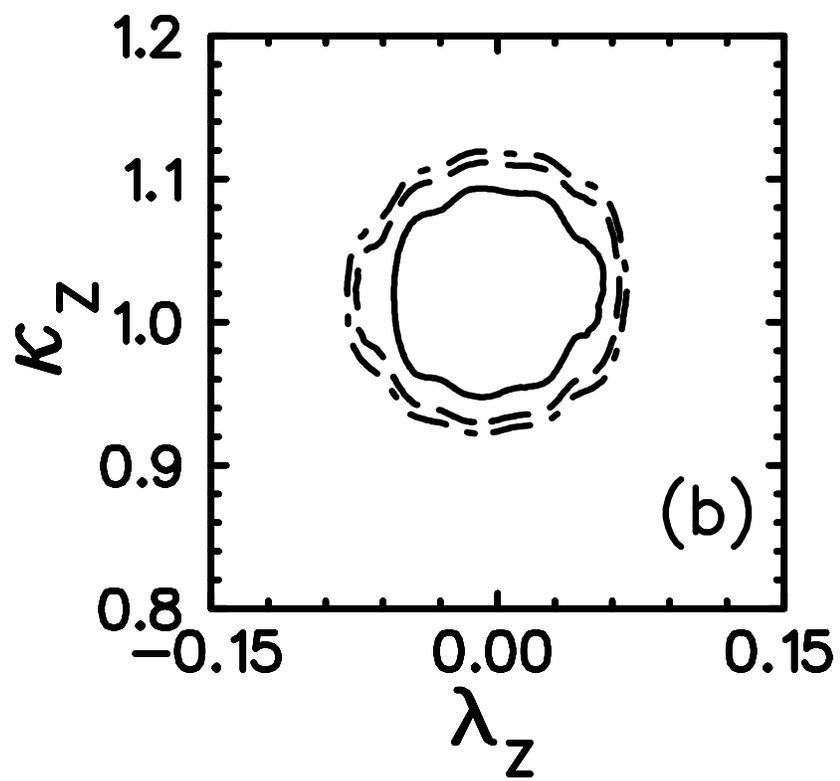
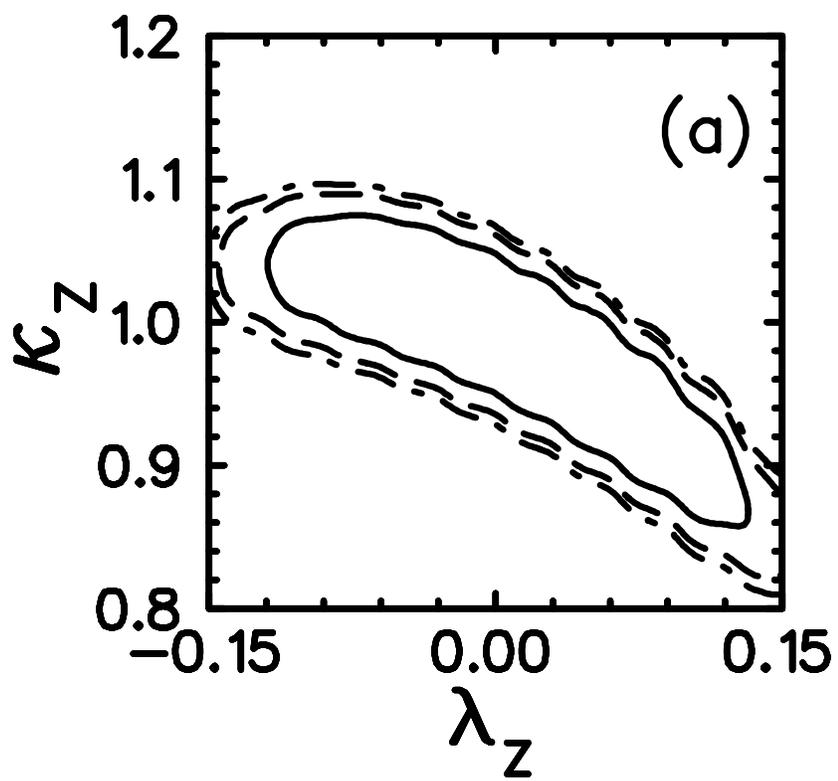
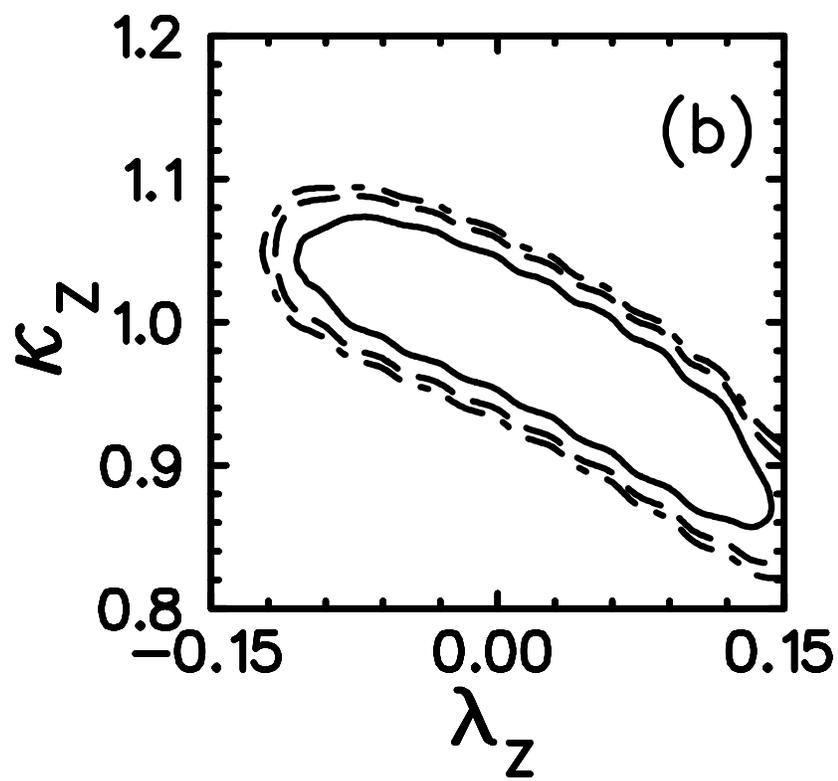
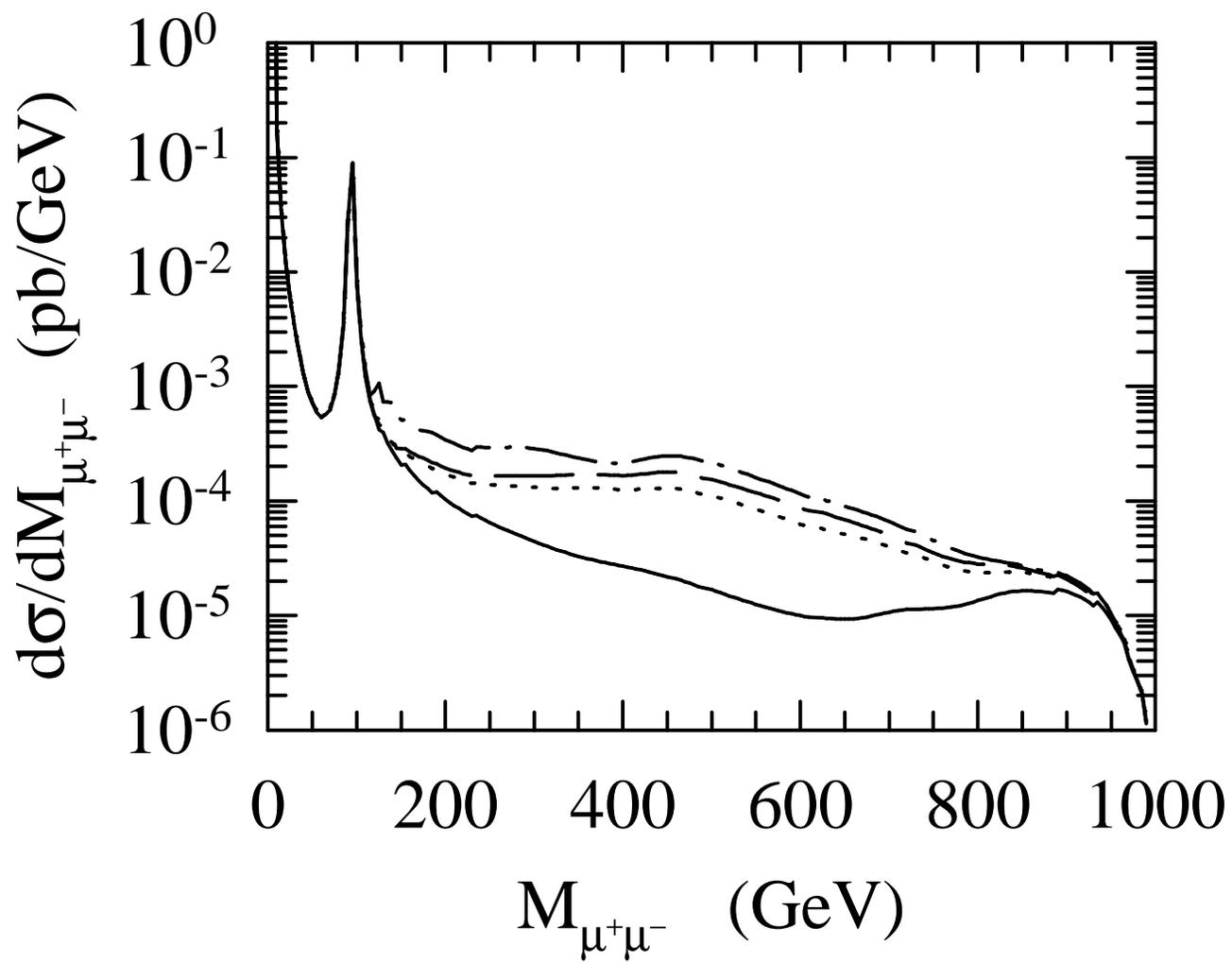
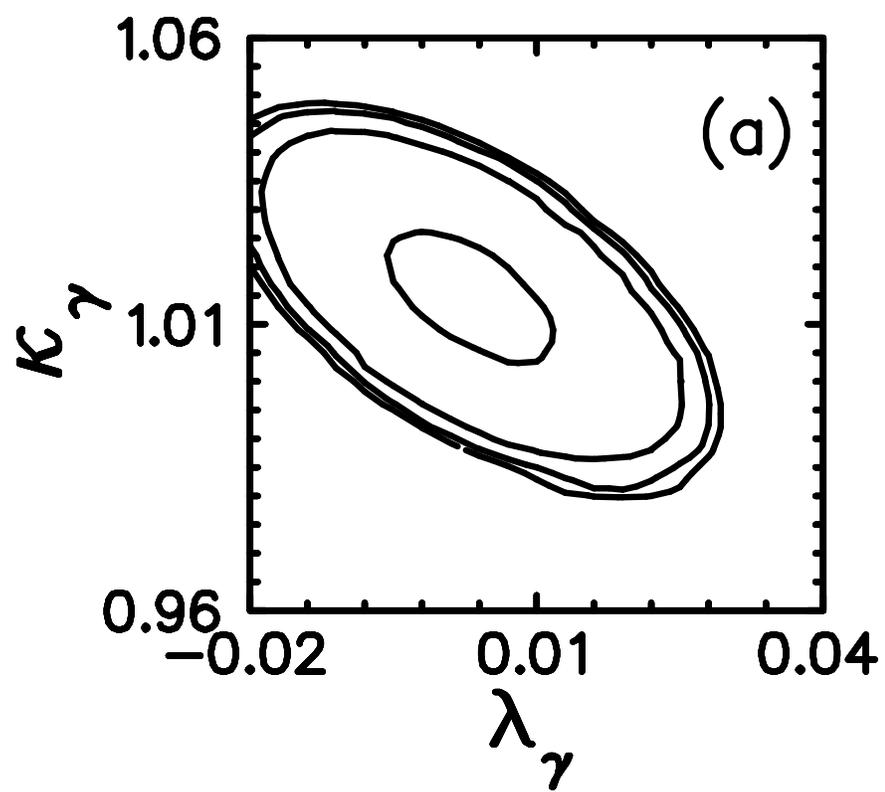
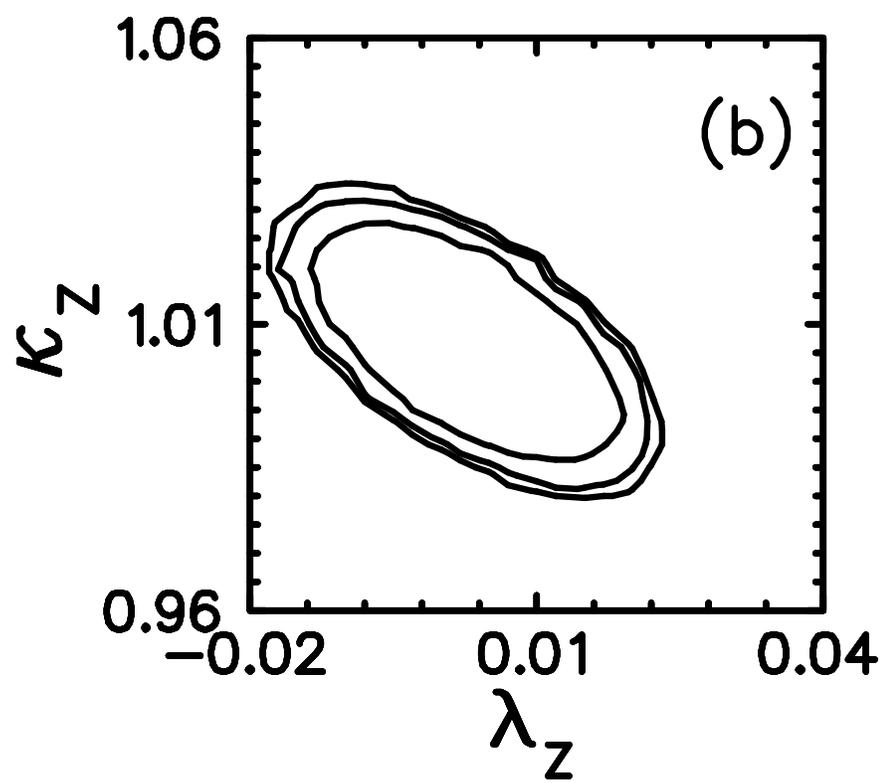
\end{document}